\def\nodes(#1){\Lambda_{#1}^0}          % the set of nodes of a graph
\def\bonds(#1){\Lambda_{#1}^1}          % the set of bonds
\def\Aij(#1,#2){A^{#1,#2}}              % the Aij
\def\Uij(#1,#2){U_{#1,#2}}              % the Uij
\def\vee{e^{\beta}-1}                   % v
\def\tensor(#1,#2,#3){#1_{#2,#3}}       % some T_{i,j}
\def\Pn{P_n(Q)}                         % Partition algebra
\def\Dg(#1){D_{#1}(Q)}                  % Diagram algebra
\def\Dd(#1){D_{#1}}                     % same, no Q
\def\Tm(#1){\mbox{\huge $\tau$}_{\!\! #1}}           % transfer matrix for
\def\ahat(#1){{\hat A}_{#1}}            % closed chain graph
\def\aha(#1){{\hat{\hat A}}_{#1}}       % the wonder graph
\def\ii(#1){\hbox{\bf I}_{#1}}          % the identity matrix
\def\ui(#1){U_{#1 \cdot}}               % the U_i
\def\Ai(#1){A^{#1 \cdot}}               % the A_i
\def\qset(#1){\{1,2,\ldots,#1\}}        % subset of the natural numbers
\def\Ss(#1){\hbox{\bf S}_{#1}}          % set of partitions
\def\Bb(#1){\hbox{\bf B}_{#1}}          % basis indexed by prop lines
\def\Hh(#1){\#^p (#1)}                  % propagating no.
\def\Ee(#1){\hbox{\bf E}_{#1}}          % idempotent
\def\Ff(#1){\hbox{\bf F}_{#1}}          % unnormalixed idempotent
\def\ep(#1){\epsilon_{#1}}              % primitive id.
\def\sym(#1){\mbox{\sf S}(#1)}          % symmetric group
\def\alt(#1){\mbox{\sf A}(#1)}          % alternating group
\def\pot(#1){\rho_{#1}}                 % potts representation
\newcommand{\beq}{\begin{equation}}
\newcommand{\eq}{\end{equation}}
\newcommand{\ba}{\begin{eqnarray}}
\newcommand{\ea}{\end{eqnarray}}
\newtheorem{theo}{Theorem}      
\newtheorem{de}{Definition}     \newtheorem{pr}{Proposition}
\newtheorem{co}{Corollary}[pr]  
\newcommand{\C}{{\bf C} \!\!\!\! I}

\def\nset(#1){ \{ #1 \}_{ \underline{n} }} %the set {#1}_{n}
\def\Sym(#1){\Sigma(#1)}                   %generic symbol for symmetric group
\def\Sy(#1){\Sigma_{#1}}                   %symmetric group irrep.
\def\sy(#1){S_{#1}}                        %another symm gp irrep.
\def\Eee(#1){{\bf E}_{\{ #1 \}_{\underline{n}}}}   %ditto for nset
\def\Sss(#1){{\bf S}_{\{ #1 \}_{\underline{n}}}}   %ditto for nset
\def\ka(#1){\kappa_{#1}}                   %maps AP_{n=#1}A to P_{n-1}
\def\SS(#1){{\cal S}_{#1}}                 %Specht/Weyl module
\def\ul(#1){_{\underline{#1}}}             % _underline #1
     
\def\eql(#1){ \begin{equation} \label{#1}
}

\def\lab(#1){\label{#1}
${}_{lab.(#1)} \;$
}
\def\prl(#1){ \begin{pr} \label{#1}
${}_{pr.(#1)} \;  \; $
}
\def\choo(#1,#2){ \left( \begin{array}{c} #1 \\ #2 \end{array} \right) }
\def\mat{ \left( \begin{array} }    \def\tam{ \end{array}  \right) }

\def\tri{\Theta^n_{n-2,0}}
\def\squ{\Theta^n_{n-3,0}}
\def\dia{\Theta^n_{n-3,1}}
\def\wgg{{}_{D_G} W_{\gamma}}
\def\whe{{}_{D_H} W_{\eta}}
\def\vgl{{}_{P_{G}} V_{\lambda}}
\def\vhm{{}_{P_{H}} V_{\mu}}
\def\res(#1,#2,#3,#4){{\mbox{Res}}^{{#1}_{#2}}_{{#3}_{#4}}}
\def\bt(#1){\mbox{\c{} {$\! #1$}}}
\def\Db{{\mbox{\bf S}}^G_{2n}}

\documentstyle[12pt]{article}
\begin{document}

\begin{center} {\bf
ON THE ALGEBRAIC APPROACH TO CUBIC LATTICE POTTS~MODELS}
\\ \vspace{.4in}
Srinandan Dasmahapatra and Paul  Martin
\footnote{Mathematics Department, City University,
Northampton Square, London EC1V 0HB, UK.} \end{center}

\begin{abstract}

We consider Diagram algebras, $\Dg(G)$ (generalized Temperley-Lieb
algebras) defined for a large
class of graphs $G$, including those of relevance for cubic lattice Potts
models, and study their structure for generic $Q$.  We find that
these algebras are too large to play the precisely analogous role
in three dimensions to that played by the Temperley-Lieb algebras
for generic $Q$ in the planar case.  We outline measures to extract the
quotient algebra that would illuminate the physics of three dimensional
Potts models.

PACS: 75.10.H
\end{abstract}

\section{ Introduction}
With the benefit of hindsight it is striking how easy it
might have
been, 15-20 years ago, to identify roots of
unity as the values of $q$ that were special for the
description of
the physics of $Q=(q+q^{-1})^2$ state Potts models in two
dimensions,
and related spin chains in one dimension.  It is the work
of a few lines
to derive these as the exceptional cases using the
Temperley-Lieb algebra introduced by Temperley and Lieb 1971 \cite{TL}
(see \cite{Mar}).
This could have been
done before many of the models were solved.  Only the interpretation of this
result might have puzzled the early `algebraic physicist'.  Of course, this is
not the way things happened.
The location of the special points is revealed in
the {\em details} of the solution of the models
\cite{Baxter}\cite{BKW}\cite{ts},
and it was only after the solution of the models
that the significance of the special points and their
relation to the cataloguing of models into universality
classes was appreciated.

In a sense, we
find ourselves heading down the same path now for three and higher
dimensional models.  There has been some very impressive
work done on models whose Boltzmann weights satisfy
the tetrahedron equations \cite{bazhanovetc}, but that is
not the route we follow here.
In \cite{marsal} it was suggested that the Diagram algebras
$\Dg(G)$ (defined below) for some sequence of graphs
$G^{(-)}=\{G^{(1)},G^{(2)},\ldots \}$ would play the role of
the Temperley-Lieb algebra for higher dimensions
\def\TL{ Temperley-Lieb }
(the \TL algebra is the sequence of Diagram algebras
with $G^{(j)}=A_j$, where $A_j$ is the $j$ node chain graph).
In this paper we determine the structure of $\Dg(G)$ for enough graphs
$G$ to show that a direct analogy with two dimensions is too simplistic
in general, and suggest a resolution.

The paper is structured in the following way.  We introduce the $Q$-state
Potts model on any lattice, and point out the relation between the
transfer matrix of the $2$-dimensional model and the Temperley-Lieb (TL)
algebra.  Since we take the algebraic route in this paper, we then state
the specific link between representation theory (the index set
for distinct irreducible representations) and physics (primary fields
in the  $2$-dimensional conformal field theory (CFT)) that we would like
to examine in the higher dimensional context.  Namely, when the index
set is {\em finite}, the corresponding CFT is {\em minimal}.  In
$2$-dimensions, the index set is finite at the special values of $Q$
called Beraha numbers, which are also the values at which the TL algebras
defined for a sequence of chain graphs of increasing length becomes
non-semisimple beyond some length.  One of our objectives in this
paper is to locate the corresponding $Q$-values at which our candidate
algebra $\Dg(G)$ becomes non-semisimple in an analogous way.

We define the Diagram algebra as a subalgebra of the Partition algebra
\cite{s2p} in the last part of this section.  The basis of the
defining representation of the Diagram algebra is taken from the set of
partitions of the nodes of two copies of a graph $G$, called
`top' and `bottom'.  Multiplication in the algebra involves
stacking one such top and bottom over
another, and keeping track of
the resulting partitions by transitivity (see figure 1).
 In section $2$, we tackle
the problem of classifying the irreducible representations of $\Dg(G)$
for generic $Q$.  This is carried out in two steps -- first by noting
the number of parts with both top and bottom nodes as above (called the
number of `propagating lines') and then by the permutations of these lines
allowed on a given graph $G$.  We do this for a large class of graphs
and in particular, for a class of graphs which we call {\em unsplitting}
(see proposition 3 and the remark following it).  We also give necessary
and sufficient conditions for a set of partitions to be a basis for these
irreducibles in proposition 6.  Using this key result, we prove in
proposition 7 that the algebras defined for a sequence of unsplitting
graphs ceases to be semi-simple for at least all integer values of $Q$.
In section 4, we apply the above results for the particular example
of an unsplitting graph that is relevant for building the transfer
matrix of the 3-dimensional Potts model.  We discuss the implication of
these results next.  The appendix lays out the preliminary steps
towards the description of the Bratteli diagram (or the inclusion
matrix) for the restriction of modules for
the generically semi-simple algebras $\Dd(H)\subset\Dd(G)$
for graphs $G,H$ and $H\subset G$.

\begin{subsection}{Basic definitions}

For any simple, unoriented graph $L$, and natural number $Q$, the partition
function of the $Q$-state Potts model \cite{Potts} on the graph $L$ is

\beq
Z(L)=\sum_{\begin{array}{c} \sigma_i\in\qset(Q) \\ \forall i\in\nodes(L)
\end{array}}
\exp\left( {\beta\sum_{(i,j)\in
\bonds(L)} \delta_{\sigma_i \sigma_j}} \right),
\eq

\noindent where $\nodes(L)$ denotes the set of nodes of $L$, and $\bonds(L)$,
the set of its edges.

Recall that for graphs $G$ and $H$ then $\;\; G\times H$ is a graph such that
\eql(01)
\nodes(G\times H)=\nodes(G)\times\nodes(H) \quad
\eq
and
\ba \cr
\bigl((i,j), (k,l)\bigr)\in&\bonds(G\times H) \quad\hbox{if}
 \cases{(i,k)\in\bonds(G) \quad
\hbox{and}\quad j=l, \quad \hbox{or}\cr (j,l)\in \bonds(H) \quad
\hbox{and}\quad i=k.\cr}
\ea
Let $\ahat(t)$ be the $t$-node closed chain graph.  Then
for example $A_l \times A_m \times \ahat(t)$ would be the cubic lattice
with periodicity in one direction.
For any $G$ the partition function
\beq
Z(G\times \ahat(t))=\hbox{Tr}\bigl( \,({\Tm(G)})^t\bigr),
\eq
where $\Tm(G)$ is the ($G$ shaped layer)
transfer matrix defined as
\beq
\Tm(G)=\prod_{i\in\nodes(G)}\Bigl( (\vee) \ii() + \sqrt Q \ui(i) \Bigr)
\prod_{(i,j)\in \bonds(G)}\Bigl( \ii() + {(\vee)\over\sqrt Q}\Uij(i,j)\Bigr).
\eq
Here
\eql(02) \ii()=\ii(Q) \otimes \ii(Q) \otimes ...  \otimes \ii(Q)
\eq
(one factor for each node of $G$, each factor a $Q \times Q$ unit matrix)
\beq
\ui(i)={1\over\sqrt Q}\bigl(\ii(Q)\otimes\ii(Q)\otimes\ldots\otimes{\underbrace
M
_{i^{th}}}\otimes\ldots\ii(Q)\bigr) \hspace{.6in} (i \in \nodes(G)),
\eq
where $M$ is the $Q\times Q$ matrix with all entries $1$,
in the $i^{th}$ position (note that writing the factors in a row
implies a total order on $\nodes(G)$ - this is
physically
misleading for general $G$ and can be chosen
arbitrarily, c.f. the two dimensional
case \cite{Baxter}) %of $\nodes(G)$ factors ,
and
\beq
\Uij(i,j)=\sqrt Q\bigl(\ii(Q)\otimes\ii(Q)\otimes\ldots\otimes{\underbrace N
_{i^{th}\otimes j^{th}}}\otimes\ldots\ii(Q)\bigr),\quad
\biggl((i,j)\in\bonds(G)\biggr)
\eq

\noindent where $N$ is the $Q^2\times Q^2$ diagonal matrix acting
on the $i^{th}$ and
$j^{th}$ subspaces (and note that $j$ is not necessarily
adjacent to $i$ in a
given ordering)
with index
set $\qset(Q)\times\qset(Q)$, and
\beq
\tensor(N, {(i,j)},{(i,j)})=\cases{1 \quad \hbox{if}\quad i=j,\cr 0 \quad
\hbox{otherwise.}\cr}
\eq
(see \cite{TL},\cite{Baxter}, \cite{MarKyo}).

Note that these matrices obey
\ba
\ui(i)^2=&\sqrt Q \ui(i)\cr
\Uij(i,j)^2=&\sqrt Q \Uij(i,j) \cr
\ui(i) \Uij(i,j) \ui(i) =&\ui(i)\cr
\Uij(i,j) \ui(i) \Uij(i,j) =& \Uij(i,j) \ea
\eql(03)
[\ui(i),\ui(j)]=[\ui(i), \Uij(j,k)]=[\Uij(i,j),\Uij(k,l)]=0, \quad i\neq j,k.
\eq

Recall that for $G=A_n$
the graph $L=G\times\ahat(t)$ is the square lattice on a cylinder, and these
matrices give a representation of the Temperley-Lieb
algebra \cite{TL}.  It is known that this representation is faithful
except at the Beraha type numbers \cite{Beraha}
 $Q=4\cos^2 {{\pi p}\over b}$ ($p,b$
integers), where it is faithful only on the unitarizable quotient \cite{J}.
Also, for other $Q$ values the number of distinct irreducible representations
in this Potts representation grows unboundedly with $n$, whereas for
$p,b$ integer it is finite and fixed by $b$ ({\em a la} primary fields
in rational conformal field theories \cite{BPZ}).
The models corresponding to these Beraha-type numbers have as massless
Euclidean field theory limits the minimal models of conformal field theory.
For $p=1$, these lattice models are in the
same universality class as the ABF models \cite{ABF}\cite{Huse}
\cite{Pas}\cite{AWK}
whose corresponding
conformal field theories belong to the unitary series of ref. \cite{FQS}
with $c=1-{6\over{b(b-1)}}$.

In this paper we address the question of what is the appropriate
abstract algebra, in the same sense as above, for arbitrary sequence
$G^{(-)}$.  In
\cite{marsal},
it has been noted that the algebra with generators and relations
simply as in equation(\ref{03})
(the Full Temperley-Lieb algebra) is too big, as the Potts representation is
then
never faithful for non-chain graphs.  Instead, we shall focus on the following
finite dimensional quotients.  In order to define these quotients, it is
useful to recall the definition of the Partition algebra $P_n = \Pn$
\cite{s2p}\cite{marsal}.

Let $\Ss(2n)$ be the set of partitions of
the set $\{1,2,\ldots,n,1',2',\ldots,n'\}$.
The $\C$-linear extension of the product defined in figure 1 on the vector
space with
basis $\Ss(2n)$ gives the Partition algebra, $\Pn$.

\def\compo{
\setlength{\unitlength}{0.00625in}%
\begin{picture}(365,415)(75,295)
\put(175,700){\makebox(0,0)[lb]{\raisebox{0pt}[0pt][0pt]{$\bullet$}}}
\put(215,700){\makebox(0,0)[lb]{\raisebox{0pt}[0pt][0pt]{$\bullet$}}}
\put(255,700){\makebox(0,0)[lb]{\raisebox{0pt}[0pt][0pt]{$\bullet$}}}
\put(295,700){\makebox(0,0)[lb]{\raisebox{0pt}[0pt][0pt]{$\bullet$}}}
\put(335,700){\makebox(0,0)[lb]{\raisebox{0pt}[0pt][0pt]{$\bullet$}}}
\put(375,700){\makebox(0,0)[lb]{\raisebox{0pt}[0pt][0pt]{$\bullet$}}}
\put(175,620){\makebox(0,0)[lb]{\raisebox{0pt}[0pt][0pt]{$\bullet$}}}
\put(215,620){\makebox(0,0)[lb]{\raisebox{0pt}[0pt][0pt]{$\bullet$}}}
\put(255,620){\makebox(0,0)[lb]{\raisebox{0pt}[0pt][0pt]{$\bullet$}}}
\put(335,620){\makebox(0,0)[lb]{\raisebox{0pt}[0pt][0pt]{$\bullet$}}}
\put(375,620){\makebox(0,0)[lb]{\raisebox{0pt}[0pt][0pt]{$\bullet$}}}
\put(295,620){\makebox(0,0)[lb]{\raisebox{0pt}[0pt][0pt]{$\bullet$}}}
\put(175,560){\makebox(0,0)[lb]{\raisebox{0pt}[0pt][0pt]{$\bullet$}}}
\put(215,560){\makebox(0,0)[lb]{\raisebox{0pt}[0pt][0pt]{$\bullet$}}}
\put(255,560){\makebox(0,0)[lb]{\raisebox{0pt}[0pt][0pt]{$\bullet$}}}
\put(295,560){\makebox(0,0)[lb]{\raisebox{0pt}[0pt][0pt]{$\bullet$}}}
\put(335,560){\makebox(0,0)[lb]{\raisebox{0pt}[0pt][0pt]{$\bullet$}}}
\put(375,560){\makebox(0,0)[lb]{\raisebox{0pt}[0pt][0pt]{$\bullet$}}}
\put(175,480){\makebox(0,0)[lb]{\raisebox{0pt}[0pt][0pt]{$\bullet$}}}
\put(215,480){\makebox(0,0)[lb]{\raisebox{0pt}[0pt][0pt]{$\bullet$}}}
\put(255,480){\makebox(0,0)[lb]{\raisebox{0pt}[0pt][0pt]{$\bullet$}}}
\put(335,480){\makebox(0,0)[lb]{\raisebox{0pt}[0pt][0pt]{$\bullet$}}}
\put(375,480){\makebox(0,0)[lb]{\raisebox{0pt}[0pt][0pt]{$\bullet$}}}
\put(295,480){\makebox(0,0)[lb]{\raisebox{0pt}[0pt][0pt]{$\bullet$}}}
\put(175,380){\makebox(0,0)[lb]{\raisebox{0pt}[0pt][0pt]{$\bullet$}}}
\put(215,380){\makebox(0,0)[lb]{\raisebox{0pt}[0pt][0pt]{$\bullet$}}}
\put(255,380){\makebox(0,0)[lb]{\raisebox{0pt}[0pt][0pt]{$\bullet$}}}
\put(295,380){\makebox(0,0)[lb]{\raisebox{0pt}[0pt][0pt]{$\bullet$}}}
\put(335,380){\makebox(0,0)[lb]{\raisebox{0pt}[0pt][0pt]{$\bullet$}}}
\put(375,380){\makebox(0,0)[lb]{\raisebox{0pt}[0pt][0pt]{$\bullet$}}}
\put(175,300){\makebox(0,0)[lb]{\raisebox{0pt}[0pt][0pt]{$\bullet$}}}
\put(215,300){\makebox(0,0)[lb]{\raisebox{0pt}[0pt][0pt]{$\bullet$}}}
\put(255,300){\makebox(0,0)[lb]{\raisebox{0pt}[0pt][0pt]{$\bullet$}}}
\put(335,300){\makebox(0,0)[lb]{\raisebox{0pt}[0pt][0pt]{$\bullet$}}}
\put(375,300){\makebox(0,0)[lb]{\raisebox{0pt}[0pt][0pt]{$\bullet$}}}
\put(295,300){\makebox(0,0)[lb]{\raisebox{0pt}[0pt][0pt]{$\bullet$}}}
\thicklines
\put(180,705){\line( 1,-2){ 42}}
\put(220,620){\line( 1, 0){ 35}}
\put(255,620){\line( 1, 2){ 43}}
\put(220,705){\line( 1,-1){ 45}}
\put(275,650){\line( 5,-6){ 25}}
\put(335,620){\line( 1, 0){ 45}}
\put(175,620){\line( 2, 1){ 30}}
\put(215,640){\line( 2, 1){ 40}}
\multiput(265,665)(0.50000,0.25000){21}{\makebox(0.4444,0.6667){\sevrm .}}
\put(290,675){\line( 2, 1){ 50}}
\put(340,560){\line( 1, 0){ 40}}
\put(300,560){\line( 0,-1){ 80}}
\put(255,560){\line( 1,-1){ 40}}
\put(305,515){\line( 1,-1){ 35}}
\put(180,565){\line( 0,-1){ 85}}
\put(180,480){\line( 1, 0){ 40}}
\put(220,300){\line(-1, 0){ 40}}
\put(180,300){\line( 2, 1){160}}
\put(220,380){\line( 1,-1){ 35}}
\put(265,335){\line( 1,-1){ 35}}
\put(175,380){\line( 3, 1){ 64.500}}
\put(240,400){\line( 3,-1){ 60}}
\multiput(300,380)(0.25000,-0.50000){21}{\makebox(0.4444,0.6667){\sevrm .}}
\put(310,360){\line( 1,-2){ 30}}
\put(160,615){\dashbox{4}(235,95){}}
\put(160,475){\dashbox{4}(235,95){}}
\put(160,295){\dashbox{4}(240,105){}}
\put(275,590){\makebox(0,0)[lb]{\raisebox{0pt}[0pt][0pt]{$\circ$}}}
\put(275,435){\makebox(0,0)[lb]{\raisebox{0pt}[0pt][0pt]{$=$}}}
\put( 75,340){\makebox(0,0)[lb]{\raisebox{0pt}[0pt][0pt]{$Q\times$}}}
\put(440,660){\makebox(0,0)[lb]{\raisebox{0pt}[0pt][0pt]{$a$}}}
\put(440,520){\makebox(0,0)[lb]{\raisebox{0pt}[0pt][0pt]{$b$}}}
\put(440,340){\makebox(0,0)[lb]{\raisebox{0pt}[0pt][0pt]{$a\circ b$}}}
\end{picture}
}

\medskip

\centerline{\compo}

\smallskip

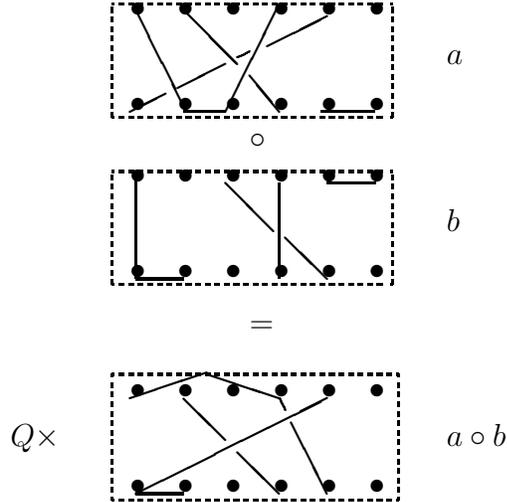
\noindent{\bf Figure 1.}{ \tenrm The top diagram is $a$,
the one in the middle $b$, and the one at the bottom is the product $a\circ b$.
Trace the connectivities from bottom to top,
and for each discarded part from the middle, pick up a factor of $Q$
to obtain $a\circ b$.}

\smallskip

The Diagram algebra, $\Dg(G)$, for a graph $G$ is defined as the subalgebra
of the Partition algebra with generators:
\ba
1=&\bigl((11')(22')\ldots(nn')\bigr),\cr
\Ai(i)=&\bigl((11')(22')\ldots(i)(i')\ldots(nn')\bigr),\quad\forall i\in
\nodes(G)\cr
\Aij(i,j)=&\bigl((11')(22')\ldots(i\ j\ i'j')\ldots(nn')\bigr),\quad\forall
(i,j)\in
\bonds(G).\cr
\ea
Note that $1^{ij}=\bigl((11')(22')\ldots(ij')\ldots(ji')\ldots(nn')\bigr)
\in \Pn$, is {\em not} in $\Dg(G)$.

The Diagram algebra may be also be thought of (visualized)
on $G \times A_k$ ($k$ large) as the
restriction of $\Pn$ to partitions achievable as connectivities
(i.e. a set of mutually non-intersecting trees
c.f. \cite{Blote})
between the
nodes of the bottom layer
(the nodes $(i,1)$ to be called
$\bt(i)$ $\forall i \in \nodes(G)$),
and those of the top layer (the nodes $(i,k)$ are to be
called $i' \forall i\in\nodes(G)$).

Note that, with $V=\C^Q$,
\beq
\pot(G):\Dg(G)\longrightarrow \hbox {End}
(V^{\otimes |\nodes(G)|}),
\eq
given by

\ba
\pot(G)(\Ai(i))=&\sqrt Q \ui(i) \cr
\pot(G)(\Aij(i,j))=&{1\over\sqrt Q} \Uij(i,j)
\ea
is a representation of the Diagram algebra called the Potts
representation (eqs. 7-8).

The Potts representation is generically faithful for $G=A_n$,
and for this reason, we here try $\Dg(G)$ as a candidate
for the appropriate
generalization of the Temperley-Lieb algebra for arbitrary graph $G$.  Note
in particular
that $\Dg(A_n)$ is isomorphic to the Temperley-Lieb algebra for any $Q$,
including non-integer values.

The partition function
$Z(L)$ may be computed working in $\Dg(G)$ instead of in the defining Potts
representation \cite{Baxter}, as in the $2$-dimensional case, where the
$\Dg(A_n)$ calculation is that of the square lattice dichromatic polynomial
\cite{BKW}\cite{Blote}.

In the two dimensional case the exceptional models may be identified
directly at the level of algebra  by finding the $Q$ values for which the
structure of the Temperley-Lieb algebra departs from the generic
semi-simple structure.
Our idea is that the departures from generic behaviour would be important
for arbitrary $G$.  The structure of $\Dg(G)$ is important
"physically," since it may be used to characterize the
spectrum of the transfer matrix, $\Tm(G)$.
Thus we proceed to analyse the structure of
$\Dg(G)$. This is already known for some $G$; in particular,
for $G=A_n$ and for $G=K_n$, the complete graph on $n$ nodes \cite{s2p}.
In this paper we consider
graphs appropriate for higher dimensional Potts models and dichromatic
polynomials, including sequences appropriate for the physically crucial
cubic lattice Potts models, and the bi-plane lattices to which
recent ideas in high $T_c$ superconductivity have drawn attention \cite{leg}.

\end{subsection}

\vskip 20pt

\section{Generic structure of $\Dg(G)$}
Mathematically, the first step in determining the structure (representation
theory) of an algebra is generally to label
the irreducible representations.
In what follows we take $n=|\nodes(G)|$.
The irreducible representations of $\Pn$ are labelled by
\[
{\cal L}_n =\{ \lambda \vdash i \; : \; i=0,1,2,...,n \}
\]
and since $\Dg(G) \subset \Pn$ all the irreducibles must be somehow contained
in the irreducibles of $\Pn$.

Consider $\Pn$ as a $\Dg(G)$ module. Clearly any $\Pn$ module is also
a $\Dg(G)$ module. Now $\Pn$ has been filtered into invariant subspaces
with bases
\[
\Bb(i) = \{ x \in \Ss(2n) \; | \; \Hh(x) \leq i \}
\]
where
$\Hh(x)$ is the number of parts of $x$ containing both primed and
unprimed nodes, called the ``propagating number'' of $x$.

If we define
\[
{\Ee(i)}^{(n)}=\prod_{j=i+1}^{n} \bigl(\frac{\Ai(j)}{Q}\bigr),
\]
then $\C$-span$(\Bb(i))=P_n \Ee(i) P_n$.
For a given $n$, we drop the superscript
$(n)$ and write $\Ee(i)$.
Note that $\Hh(\Ee(i))=i$ and for $a,b\in\Ss(2n)$,
\[
\Hh(ab)
\leq min(\Hh(a),\Hh(b))
,
\]
and we ignore elements of $\C$ in evaluating
$\Hh(z)\; \forall z\in\Pn$.
Thus
\[
P_n[i]=P_n \Ee(i) P_n / P_n \Ee(i-1) P_n
\]
is a $\Pn$ module with basis $\Bb(i) \setminus \Bb(i-1)$.
Note that in the diagrammatic realization of the left action of
$\Dg(G)$ on $P_n[i]$ the ``bottom'' of each $x \in \Bb(i)
\setminus\Bb(i-1)$
(i.e. the connectivities of the unprimed nodes of any $x\in\Ss(2n)$)
remains
unchanged. That is, all elements with the same bottom form a
submodule.

For example $\Delta_i:=P_n \Ee(i)$ (mod. $P_n \Ee(i-1) P_n$)
is one of the left
$P_n$ submodules of $P_n[i]$, and
$P_n[i]$ may be decomposed into submodules
all of which are isomorphic.
Note that $\Delta_i$ has a basis the set of partitions
which have each unprimed node in a different
part, the last $n-i$ nodes singletons
(i.e. in parts on their own),
the others connected to primed nodes.

In fact as a left $\Dg(G)$ module $\Delta_i$ breaks as
$\Dg(G) \Ee(i) \oplus R_i$ where $R_i$ is either empty or a direct sum
of one dimensional modules (see Appendix), so we need only focus on
$\Dd(G) \Ee(i)$  mod $\Dd(G) \Ee(i-1) \Dd(G)$.

The final piece of the jigsaw for $\Pn$ is to note that
$P_n \Ee(i)$ is a projective right $\sym(i)$ module (i.e. a direct summand
of a direct sum of copies of the regular representation of
the symmetric group \cite{cohn})
where the action is to permute the first $i$ (unprimed) nodes.
For example, see figure 2.

\medskip
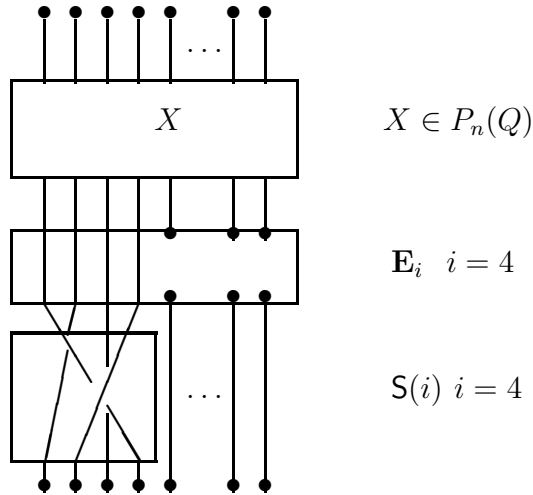
\centerline{
\setlength{\unitlength}{0.00825in}%
\begin{picture}(280,306)(140,480)
\thicklines
\put(160,780){\line( 0,-1){ 40}}
\put(180,780){\line( 0,-1){ 40}}
\put(200,780){\line( 0,-1){ 40}}
\put(220,780){\line( 0,-1){ 40}}
\put(280,780){\line( 0,-1){ 40}}
\put(300,780){\line( 0,-1){ 40}}
\put(140,680){\framebox(180,60){}}
\put(160,680){\line( 0,-1){ 40}}
\put(180,680){\line( 0,-1){ 40}}
\put(200,680){\line( 0,-1){ 40}}
\put(220,680){\line( 0,-1){ 40}}
\put(280,680){\line( 0,-1){ 40}}
\put(300,680){\line( 0,-1){ 40}}
\put(160,600){\line( 3,-5){ 30}}
\put(200,535){\line( 3,-5){ 20.735}}
\put(180,500){\line( 2, 5){ 40}}
\put(180,600){\line(-1,-4){  5}}
\put(175,570){\line(-1,-5){ 14.039}}
\put(200,600){\line( 0,-1){ 40}}
\put(200,530){\line( 0,-1){ 30}}
\put(160,500){\line( 0,-1){ 20}}
\put(180,500){\line( 0,-1){ 20}}
\put(200,500){\makebox(0.4444,0.6667){\tenrm .}}
\put(200,500){\line( 0,-1){ 20}}
\put(220,500){\line( 0,-1){ 20}}
\put(160,640){\line( 0,-1){ 40}}
\put(180,640){\line( 0,-1){ 40}}
\put(200,640){\line( 0,-1){ 40}}
\put(220,640){\line( 0,-1){ 40}}
\put(140,600){\framebox(180,45){}}
\put(140,500){\framebox(90,80){}}
\put(240,780){\line( 0,-1){ 40}}
\put(240,680){\line( 0,-1){ 35}}
\put(240,600){\line( 0,-1){120}}
\put(280,600){\line( 0,-1){120}}
\put(300,600){\line( 0,-1){120}}
\put(155,780){\makebox(0,0)[lb]{\raisebox{0pt}[0pt][0pt]{$\bullet$}}}
\put(175,780){\makebox(0,0)[lb]{\raisebox{0pt}[0pt][0pt]{$\bullet$}}}
\put(195,780){\makebox(0,0)[lb]{\raisebox{0pt}[0pt][0pt]{$\bullet$}}}
\put(215,780){\makebox(0,0)[lb]{\raisebox{0pt}[0pt][0pt]{$\bullet$}}}
\put(275,780){\makebox(0,0)[lb]{\raisebox{0pt}[0pt][0pt]{$\bullet$}}}
\put(295,780){\makebox(0,0)[lb]{\raisebox{0pt}[0pt][0pt]{$\bullet$}}}
\put(275,640){\makebox(0,0)[lb]{\raisebox{0pt}[0pt][0pt]{$\bullet$}}}
\put(295,640){\makebox(0,0)[lb]{\raisebox{0pt}[0pt][0pt]{$\bullet$}}}
\put(275,600){\makebox(0,0)[lb]{\raisebox{0pt}[0pt][0pt]{$\bullet$}}}
\put(295,600){\makebox(0,0)[lb]{\raisebox{0pt}[0pt][0pt]{$\bullet$}}}
\put(155,480){\makebox(0,0)[lb]{\raisebox{0pt}[0pt][0pt]{$\bullet$}}}
\put(175,480){\makebox(0,0)[lb]{\raisebox{0pt}[0pt][0pt]{$\bullet$}}}
\put(195,480){\makebox(0,0)[lb]{\raisebox{0pt}[0pt][0pt]{$\bullet$}}}
\put(215,480){\makebox(0,0)[lb]{\raisebox{0pt}[0pt][0pt]{$\bullet$}}}
\put(375,710){\makebox(0,0)[lb]{\raisebox{0pt}[0pt][0pt]{$ X \in \Pn$}}}
\put(230,710){\makebox(0,0)[lb]{\raisebox{0pt}[0pt][0pt]{$ X$}}}
\put(380,620){\makebox(0,0)[lb]{\raisebox{0pt}[0pt][0pt]{$ \Ee(i)$}}}
\put(380,540){\makebox(0,0)[lb]{\raisebox{0pt}[0pt][0pt]{$ \sym(i)$}}}
\put(235,780){\makebox(0,0)[lb]{\raisebox{0pt}[0pt][0pt]{$\bullet$}}}
\put(235,640){\makebox(0,0)[lb]{\raisebox{0pt}[0pt][0pt]{$\bullet$}}}
\put(235,600){\makebox(0,0)[lb]{\raisebox{0pt}[0pt][0pt]{$\bullet$}}}
\put(235,480){\makebox(0,0)[lb]{\raisebox{0pt}[0pt][0pt]{$\bullet$}}}
\put(275,480){\makebox(0,0)[lb]{\raisebox{0pt}[0pt][0pt]{$\bullet$}}}
\put(295,480){\makebox(0,0)[lb]{\raisebox{0pt}[0pt][0pt]{$\bullet$}}}
\put(250,760){\makebox(0,0)[lb]{\raisebox{0pt}[0pt][0pt]{$ \ldots$}}}
\put(250,540){\makebox(0,0)[lb]{\raisebox{0pt}[0pt][0pt]{$ \ldots$}}}
\put(415,620){\makebox(0,0)[lb]{\raisebox{0pt}[0pt][0pt]{$ i=4$}}}
\put(420,540){\makebox(0,0)[lb]{\raisebox{0pt}[0pt][0pt]{$ i=4$}}}
\end{picture}
}

\medskip

\noindent{\bf Figure 2} {\tenrm The right action of the
permutation group (depicted in the box) is, diagrammatically,
the action from below.}

\medskip

\noindent Thus $P_n \Ee(i)$ (mod) breaks up into simple modules indexed by
$\lambda \vdash i$ (from $\sym(i)$ representation theory \cite{ham}).

For $\Dg(G)$, however, the picture is more complicated,
since $\Dd(G)\Ee(i)$ is not always closed under the
right action of $\sym(i)$.  For example, whereas
$\Pn \Ee(n)\cong\sym(n)$ modulo $\Pn\Ee(n-1)\Pn$,
we have $\Dd(G) \Ee(n) = \C \Ee(n)=\C\cdot 1$ mod.
$\Dd(G) \Ee(n-1) \Dd(G)$
for any $G$.
To see this note that with $n$ propagating lines from
bottom to
top of $G \times A_k$ ($k$ large) there is only one possibility,
as depicted in figure 3.

\medskip

\centerline{
\setlength{\unitlength}{0.0100in}%
\begin{picture}(120,320)(280,400)
\thicklines
\put(280,680){\line( 2, 1){ 40}}
\put(320,700){\line( 1, 0){ 40}}
\put(360,700){\line( 2,-1){ 40}}
\put(400,680){\line(-1,-1){ 20}}
\put(380,660){\line(-1, 0){ 80}}
\put(300,660){\line(-1, 1){ 20}}
\put(280,620){\line( 2, 1){ 40}}
\put(320,640){\line( 1, 0){ 40}}
\put(360,640){\line( 2,-1){ 40}}
\put(400,620){\line(-1,-1){ 20}}
\put(380,600){\line(-1, 0){ 80}}
\put(300,600){\line(-1, 1){ 20}}
\put(280,560){\line( 2, 1){ 40}}
\put(320,580){\line( 1, 0){ 40}}
\put(360,580){\line( 2,-1){ 40}}
\put(400,560){\line(-1,-1){ 20}}
\put(380,540){\line(-1, 0){ 80}}
\put(300,540){\line(-1, 1){ 20}}
\put(280,500){\line( 2, 1){ 40}}
\put(320,520){\line( 1, 0){ 40}}
\put(360,520){\line( 2,-1){ 40}}
\put(400,500){\line(-1,-1){ 20}}
\put(380,480){\line(-1, 0){ 80}}
\put(300,480){\line(-1, 1){ 20}}
\put(280,440){\line( 2, 1){ 40}}
\put(320,460){\line( 1, 0){ 40}}
\put(360,460){\line( 2,-1){ 40}}
\put(400,440){\line(-1,-1){ 20}}
\put(380,420){\line(-1, 0){ 80}}
\put(300,420){\line(-1, 1){ 20}}
\put(280,420){\framebox(2,280){}}
\put(300,400){\framebox(2,280){}}
\put(320,440){\framebox(2,280){}}
\put(355,440){\framebox(2,280){}}
\put(375,400){\framebox(2,280){}}
\put(395,420){\framebox(2,280){}}
\end{picture}
}

\medskip

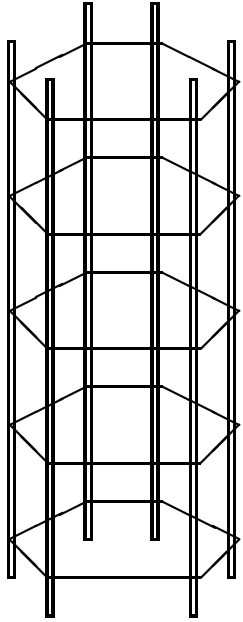
\noindent{\bf Figure 3} {\tenrm There is no space for
lateral motion if all of the nodes of $G$ (the hexagon)
are propagating. The propagating lines are drawn with double lines.}

\medskip

\noindent Our problem is thus reduced to determining the maximum subgroup
$H_G^i\subset\sym(i)$ for which
$\Dd(G) \Ee(i)$ (mod $\Dd(G)\Ee(i-1)\Dd(G)$)
is a right module.  In general, for $i<n$, the
situation depends on $G$.

Before actually determining $H_G^i$, let us first explicitly
construct words in the algebra that would implement
the group
action.  As is clear from figure 3, one or more nodes
of $G$ need to be disconnected to allow for walks on
$G\times A_k$ to realize any permutations of the nodes (except for the
identity permutation as in figure 3).
Also, since there is no unique, or natural ordering
of the nodes of $G$, we need to determine whether $H_G^i$
depends on the choice of the nodes
disconnected by $\Ee(i)$.

For any subset, $s$ of $\qset(n)$, let $p_s$ be the set difference
$\qset(n)\setminus s$, and $\Ee(\{p_s\})=\prod_{j\in s} (\Ai(j)/Q)$.
For example,
for $\underline{i}=\{i+1,i+2,\ldots,n\}$,
\beq
p_{\underline{i}}=\qset(i) \quad\hbox{and}\quad \Ee(\{p_{\underline{i}}\})=
\prod_{j=i+1}^{n} \bigl({\Ai(j)\over Q}\bigr)=\Ee(i).
\eq

\begin{de}
The partition basis of $\Dg(G)$, denoted by $\Db$ is
$\Ss(2n)\cap\Dd(G)$.  Also, set
${\Bb(i)}^G:=\Bb(i)\cap \Dd(G)$.
\end{de}

\begin{de}
Let $s,t\subset\qset(n)$,  s.t. $|s|=|t|$.  Then
\eql(fist)
\Phi_s^t:=\{\varphi_s^t | \varphi_s^t=\Ee(\{p_t\})X\Ee(\{p_s\}),\;
\forall X\in \Db \;
s.t.\; \Hh(\Ee(\{p_t\})X\Ee(\{p_s\}))=|p_s|\}.
\eq
\end{de}

These elements of (\ref{fist}) may be interpreted as
bijections, $\varphi_s^t:p_s\rightarrow p_t$.
Note, in particular, that $\Phi_s^s\subseteq \sym(p_s)
\Ee({\{p_s\}})$.

Let $\delta=s\bigtriangleup t=(s\setminus t)\cup
(t\setminus s)$, the symmetric difference of sets $s,t$,
with $|\delta|=2d$, i.e., $d$ elements of the $n-i$
elements of $s$ are distinct from those of $t$.  Then $\exists$ partitions
of $\delta$ of shape $2^d$, i.e. of the form
\beq
\bigl(
(\delta_1\delta^{*}_1)(\delta_2\delta^{*}_2)\cdots(\delta_d\delta^{*}_d)
\bigr),
\eq
where the unstarred nodes of $\delta \in s$ and the starred ones in $t$.
\def\xmui(#1){x^{(\mu_i)}_{#1}}
Consider chain subgraphs, $A^{(i)}_{\mu_i}, i=1,\ldots, d$ of the connected
graph $G$, with nodes labelled $\xmui(j), j=1,2\ldots,\mu_i$
such that the first node of
$A^{(i)}_{\mu_i}$ is $\xmui(1)=\delta_i$ and the $\mu_i^{th}$,
$\xmui(\mu_i)=\delta^{*}_i$.
Let us construct words $\omega_{A^{(i)}_{\mu_i}}$ of the form
\eql(move)
\omega_{A^{(i)}_{\mu_i}}=\prod_{j=0}^{\mu_i-2} \Bigl(
\Ai({\xmui(\mu_i-j)})\;
\Aij({\xmui(\mu_i-j)},{\xmui(\mu_i-j-1)})\Bigr)
\Ai({\xmui(1)})
\eq
s.t. $\omega_{A^{(i)}_{\mu_i}}$ achieves the connectivity which differs
from the unit in
\beq
\bigl(\cdots
(\bt({\xmui(1)}))(\bt({\xmui(2)}){\xmui(1)}')
(\bt({\xmui(3)}){\xmui(2)}')\cdots(\bt({\xmui(\mu_i)})
{\xmui(\mu_i-1)}')({\xmui(\mu_i)}')
\cdots\bigr)\in \Dg(G)
\eq
on the sublattice $A^{(i)}_{\mu_i}\times A_k$, ($k>\mu_i$),
where (as before) $(\xmui(j),1)=\bt({\xmui(j)})$ and
$(\xmui(j),k)={\xmui(j)}'$. (See figure 4.)

\def\shift{
\setlength{\unitlength}{0.00625in}%
\begin{picture}(240,189)(160,560)
\thicklines
\put(180,740){\line( 0,-1){175}}
\put(180,565){\line( 0,-1){  5}}
\put(200,740){\line( 0,-1){160}}
\put(200,580){\line( 1, 0){ 20}}
\put(220,580){\line( 0,-1){ 20}}
\put(220,740){\line( 0,-1){140}}
\put(220,600){\line( 1, 0){ 20}}
\put(240,600){\line( 0,-1){ 40}}
\put(240,560){\makebox(0.4444,0.6667){\tenrm .}}
\put(240,740){\line( 0,-1){120}}
\put(240,620){\line( 1, 0){ 20}}
\put(260,620){\line( 0,-1){ 60}}
\put(260,740){\line( 0,-1){100}}
\put(260,640){\line( 1, 0){ 20}}
\put(280,640){\line( 0,-1){ 80}}
\put(280,740){\line( 0,-1){ 80}}
\put(280,660){\line( 1, 0){ 20}}
\put(300,660){\line( 0,-1){100}}
\put(300,740){\line( 0,-1){ 60}}
\put(300,680){\line( 1, 0){ 20}}
\put(320,680){\line( 0,-1){120}}
\put(320,740){\line( 0,-1){ 40}}
\put(320,700){\line( 1, 0){ 20}}
\put(340,700){\line( 0,-1){140}}
\put(340,740){\line( 0,-1){ 20}}
\put(340,720){\line( 1, 0){ 20}}
\put(360,720){\line( 0,-1){160}}
\put(380,740){\line( 0,-1){180}}
\put(400,740){\line( 0,-1){180}}
\put(160,740){\line( 0,-1){180}}
\put(355,740){\makebox(0,0)[lb]{\raisebox{0pt}[0pt][0pt]{$\delta_i^{*}$}}}
\put(195,560){\makebox(0,0)[lb]{\raisebox{0pt}[0pt][0pt]{$\delta_i$}}}
\end{picture}
}
\medskip

\centerline{\shift}

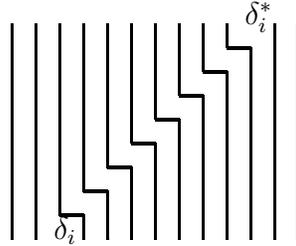
\noindent {\bf Figure 4} {\tenrm The element of the algebra shifting the
``hole''
from $\delta_i$ to $\delta_i^{*}$.  Note the
minimum height required to achieve
this connectivity is of the order of
the distance $|\delta_i - \delta_i^{*}|$.}

\smallskip

It is useful to view the element of the algebra as one that pushes a ``hole"
from its location in $s$ to one in
$t$.
The equivalence relation that defines the algebra $\Dg(G)\subset\Pn$
implies that
\eql(phi)
\omega_{\{\mu_i\}}=\prod_{i=1}^d
\omega_{A_{\mu_i}}\in\Phi_s^t,
\eq
independent of the choice of graphs $A^{(i)}_{\mu_i}$ connecting the nodes of
$\delta$.  Thus,

\begin{pr}
\eql(did)
\Dd(G)\Ee(\{p_s\})\Dd(G) = \Dd(G)\Ee(i)\Dd(G),\, \,\forall s\subset\qset(n),\,
\mbox{s.t.}|s|=n-i.
\eq
\end{pr}

The different choices of pairing the starred and unstarred
nodes of $\delta$ give
different bijections $\varphi_s^t\in\Phi_s^t$.  Let $H_G^{p_s}:=\Phi_s^s$.
We then have

\begin{pr}

For any fixed element ${\varphi}_t^s\in\Phi_t^s$,
$\;{\varphi}_t^s\Phi_s^t=H_G^{p_s}$ and $H_G^{p_s}=H_G^{p_t}$ if
$|s|=|t|$.

\end{pr}

\noindent{\em Proof}:
For sets $s,t,r$ of the same cardinality,
these `bijections' obey
\beq
\rho_t^r\circ\varphi_s^t\in\Phi_s^r \quad \forall\rho_t^r\in\Phi_t^r
\quad \hbox{and}\quad
\varphi_s^t\in\Phi_s^t.
\eq
Therefore,
\beq
\Phi_t^r\Phi_s^t\subseteq\Phi_s^r\Rightarrow|\Phi_t^r|\leq
|\Phi_s^r|\,
\mbox{and}  \,
|\Phi_s^t|\leq|\Phi_s^r| \Rightarrow|\Phi_s^r|=|\Phi_t^u|
\eq
for any $r,s,t,u\in\qset(n)$ with $|r|=|s|=|t|=|u|$.

In particular,
$|\Phi_t^s|=|\Phi_s^s|$ and
${\varphi}_t^s\Phi_s^t\subseteq\Phi_s^s$
$\Rightarrow\, {\varphi}_t^s\Phi_s^t=\Phi_s^s$, for
a fixed ${\varphi}_t^s\Phi_t^s$.
Also, $\rho_s^t \Phi_s^s \varphi_t^s=\Phi_t^t$.  This
implies that $\Phi_s^s\cong\Phi_t^t$ and depends only on
the cardinality, $|p_s|$.

{\hfill $\Box$}

\begin{co}
\eql(ron)
\Dd(G) \Ee(\{p_s\}) \cong \Dd(G) \Ee(i) \;\; \mbox{and}\;\;
\Ee(i) \Dd(G) \Ee(i)  \cong  \Ee(i)\otimes H^i_G
\hspace{.3in} mod.\;\; \Dd(G) \Ee(i-1) \Dd(G)
\eq
where $H^{i-1}_G$ is
no smaller than the maximal subgroup of $\sym(i-1)$
contained in $H^i_G$ (so that in particular if any
$H^i_G=\sym(i)$ then $H^j_G=\sym(j) \; \forall j<i$).
\end{co}

Hence,

\begin{theo}
Let $\Gamma^i_G,\;\Gamma_G$ be index sets for irreducible representations of
$H_G^i$ and $\Dg(G)$ respectively.
Then
\[
\Gamma_G=\cup_i \Gamma^i_G.
\]

\end{theo}

\begin{subsection} {How to compute $H_G^i$.}

For an arbitrary graph $G$, $|\nodes(G)|=n$, and $\alpha\in
\nodes(G)$, consider closed chain subgraphs,
$\ahat(p+1)\subset G$ with
$\alpha\in\nodes({\ahat(p+1)})$.  Setting $d=1$,
$\xmui(1)=\xmui(\mu_i)$ and $\mu_i=p+2$ in (\ref{phi}),
we note $\Phi_{\{\alpha\}}^{\{\alpha\}}$
contains ${\sf Z}_p$.
By pushing the hole around, by (\ref{move}), so as to
lie on other closed chain subgraphs, the set of
these $p$-cycles generate $H_G^{n-1}$.

\def\twist{
\setlength{\unitlength}{0.00850in}%
\begin{picture}(60,329)(300,435)
\thicklines
\multiput(300,680)(6.66667,6.66667){4}{\makebox(0.4444,0.6667){\tenrm .}}
\multiput(340,680)(6.66667,6.66667){4}{\makebox(0.4444,0.6667){\tenrm .}}
\multiput(340,720)(6.66667,6.66667){4}{\makebox(0.4444,0.6667){\tenrm .}}
\multiput(300,720)(6.66667,6.66667){4}{\makebox(0.4444,0.6667){\tenrm .}}
\multiput(300,600)(6.66667,6.66667){4}{\makebox(0.4444,0.6667){\tenrm .}}
\multiput(340,600)(6.66667,6.66667){4}{\makebox(0.4444,0.6667){\tenrm .}}
\multiput(300,560)(6.66667,6.66667){4}{\makebox(0.4444,0.6667){\tenrm .}}
\multiput(340,560)(6.66667,6.66667){4}{\makebox(0.4444,0.6667){\tenrm .}}
\multiput(320,740)(0.00000,-9.09091){23}{\makebox(0.4444,0.6667){\tenrm .}}
\multiput(300,720)(0.00000,-9.09091){23}{\makebox(0.4444,0.6667){\tenrm .}}
\multiput(300,520)(6.66667,6.66667){4}{\makebox(0.4444,0.6667){\tenrm .}}
\multiput(300,640)(6.66667,6.66667){4}{\makebox(0.4444,0.6667){\tenrm .}}
\multiput(340,720)(0.00000,-9.09091){23}{\makebox(0.4444,0.6667){\tenrm .}}
\multiput(340,520)(6.66667,6.66667){4}{\makebox(0.4444,0.6667){\tenrm .}}
\multiput(360,740)(0.00000,-9.09091){23}{\makebox(0.4444,0.6667){\tenrm .}}
\multiput(340,640)(6.66667,6.66667){4}{\makebox(0.4444,0.6667){\tenrm .}}
\multiput(320,540)(0.00000,-10.00000){5}{\makebox(0.4444,0.6667){\tenrm .}}
\multiput(360,540)(0.00000,-10.00000){5}{\makebox(0.4444,0.6667){\tenrm .}}
\multiput(300,520)(0.00000,-10.00000){5}{\makebox(0.4444,0.6667){\tenrm .}}
\multiput(300,480)(6.66667,6.66667){4}{\makebox(0.4444,0.6667){\tenrm .}}
\multiput(340,520)(0.00000,-10.00000){5}{\makebox(0.4444,0.6667){\tenrm .}}
\multiput(340,480)(6.66667,6.66667){4}{\makebox(0.4444,0.6667){\tenrm .}}
\put(300,750){\line( 0,-1){150}}
\put(300,600){\line( 1, 0){ 40}}
\put(340,600){\line( 0,-1){160}}
\put(320,760){\line( 0,-1){155}}
\put(320,595){\line( 0,-1){ 15}}
\put(320,580){\line(-1,-1){ 20}}
\put(300,560){\line( 0,-1){120}}
\put(340,750){\line( 0,-1){110}}
\put(340,640){\line( 1, 1){ 20}}
\put(360,660){\line( 0,-1){120}}
\put(360,540){\line(-1, 0){ 15}}
\multiput(300,720)(10.00000,0.00000){5}{\makebox(0.4444,0.6667){\tenrm .}}
\multiput(320,700)(10.00000,0.00000){5}{\makebox(0.4444,0.6667){\tenrm .}}
\multiput(300,680)(10.00000,0.00000){5}{\makebox(0.4444,0.6667){\tenrm .}}
\multiput(320,660)(10.00000,0.00000){5}{\makebox(0.4444,0.6667){\tenrm .}}
\multiput(320,620)(10.00000,0.00000){5}{\makebox(0.4444,0.6667){\tenrm .}}
\multiput(320,580)(10.00000,0.00000){5}{\makebox(0.4444,0.6667){\tenrm .}}
\multiput(300,520)(10.00000,0.00000){5}{\makebox(0.4444,0.6667){\tenrm .}}
\multiput(300,480)(10.00000,0.00000){5}{\makebox(0.4444,0.6667){\tenrm .}}
\multiput(320,500)(10.00000,0.00000){5}{\makebox(0.4444,0.6667){\tenrm .}}
\multiput(300,560)(10.00000,0.00000){5}{\makebox(0.4444,0.6667){\tenrm .}}
\multiput(300,640)(10.00000,0.00000){5}{\makebox(0.4444,0.6667){\tenrm .}}
\multiput(320,740)(10.00000,0.00000){5}{\makebox(0.4444,0.6667){\tenrm .}}
\put(335,540){\line(-1, 0){ 15}}
\put(320,540){\line( 0,-1){ 80}}
\put(300,745){\makebox(0,0)[lb]{\raisebox{0pt}[0pt][0pt]{$ A$}}}
\put(320,755){\makebox(0,0)[lb]{\raisebox{0pt}[0pt][0pt]{$ B$}}}
\put(340,745){\makebox(0,0)[lb]{\raisebox{0pt}[0pt][0pt]{$ C$}}}
\put(300,435){\makebox(0,0)[lb]{\raisebox{0pt}[0pt][0pt]{$ B$}}}
\put(340,435){\makebox(0,0)[lb]{\raisebox{0pt}[0pt][0pt]{$ A$}}}
\put(320,455){\makebox(0,0)[lb]{\raisebox{0pt}[0pt][0pt]{$ C$}}}
\end{picture}
}
\medskip
\centerline{\twist}
\medskip

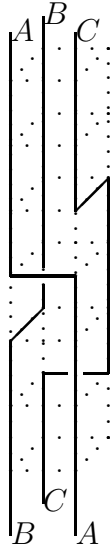
\noindent {\bf Figure 5}  {\tenrm An example of a 3-cycle (BAC), with
$G=A_2\times A_2$.
If we label the nodes such that at the bottom layer,
A,B and C are drawn through 1,2 and 3 respectively,
with the ``hole" at 4, $p_{\{4\}}=\{$1,2,3$\}$
and the connectivity drawn is ((\c{1}2$'$)(\c{2}3$'$)(\c{3}1$'$))$
\in\Phi_{\{4\}}^{\{4\}}$.}

\smallskip

\def\gx{G_x^{\times}}
\def\gxy{{G^{\times}}_{(x,y)}}
For any graph $G$, let $\gx$ denote the
graph obtained by
removing the node $x\in\nodes(G)$ and the bonds connected
to it, i.e., $\nodes(\gx)=\qset(n)\setminus \{x\}$
and $\bonds(\gx)=\bonds(G)\setminus
\{(i,x)|(i,x)\in\bonds(G)\}$.  Let $\gxy$
denote the graph obtained by removing the bond $(x,y)$,
i.e., $\nodes(\gxy)=\nodes(G)$ and
$\bonds(G)\setminus\bonds(\gxy)=\{(x,y)\}$.
Recall,
$\Ee(n-1)\Dd(G)\Ee(n-1)\cong \Ee(n-1)\otimes H_G^{n-1}$.
Then, we have

\begin{pr}

Let $|\nodes(G_i)|=n_i\forall i$.

\noindent i) If $\gxy=G_1\sqcup G_2$, then
$H_G^{n-1}=H_{G_1}^{n_1-1}\times H_{G_2}^{n_2-1}$ where the
2 factors act on the $n_1-1$ and $n_2-1$ nodes
in $G_1$ and $G_2$ respectively.

\noindent ii) If $\gx=G'_1\sqcup G'_2 \sqcup \cdots$, then
$\exists G_i\supset G'_i\, \forall i$ s.t. ${\cap}_i
\nodes(G_i)=\{x\}$ and $\cap_i \bonds(G_i)=\{\emptyset\}$,
s.t.
$H_G^{n-1}=H_{G_1}^{n_1-1} \times
H_{G_2}^{n_2-1}\times\cdots$, where $H_{G_i}^{n_i-1}$ acts
on $G_i$ only.

\end{pr}

\noindent {\bf Remark.}
Let us call graphs $G$ that do not
decompose in the sense of the previous proposition
{\em unsplitting}.  A simple example of an unsplitting
graph is the closed chain graph $\ahat(n)$.  We shall
consider other examples below.

\def\figtri{\setlength{\unitlength}{0.00625in}%
\begin{picture}(175,189)(150,565)
\thicklines
\put(160,660){\line( 1, 3){ 20}}
\put(180,720){\line( 2, 1){ 40}}
\put(220,740){\line( 1, 0){ 40}}
\put(260,740){\line( 5,-3){ 34.559}}
\put(295,720){\line( 3,-5){ 26.471}}
\put(320,675){\line( 0, 1){  5}}
\put(320,680){\line( 0,-1){ 40}}
\put(320,640){\line(-1,-2){ 20}}
\put(300,600){\line(-2,-1){ 40}}
\put(260,580){\line(-1, 0){ 40}}
\put(220,580){\line(-2, 1){ 40}}
\put(180,600){\line(-1, 3){ 20}}
\put(180,715){\line( 0,-1){115}}
\put(150,665){\makebox(0,0)[lb]{\raisebox{0pt}[0pt][0pt]{\tenrm 1}}}
\put(170,720){\makebox(0,0)[lb]{\raisebox{0pt}[0pt][0pt]{\tenrm 2}}}
\put(210,740){\makebox(0,0)[lb]{\raisebox{0pt}[0pt][0pt]{\tenrm 3}}}
\put(265,745){\makebox(0,0)[lb]{\raisebox{0pt}[0pt][0pt]{\tenrm 4}}}
\put(300,720){\makebox(0,0)[lb]{\raisebox{0pt}[0pt][0pt]{\tenrm 5}}}
\put(325,680){\makebox(0,0)[lb]{\raisebox{0pt}[0pt][0pt]{\tenrm 6}}}
\put(210,570){\makebox(0,0)[lb]{\raisebox{0pt}[0pt][0pt]{\tenrm n-1}}}
\put(255,565){\makebox(0,0)[lb]{\raisebox{0pt}[0pt][0pt]{\tenrm n-2}}}
\put(170,595){\makebox(0,0)[lb]{\raisebox{0pt}[0pt][0pt]{\tenrm n}}}
\end{picture}}

\def\figsqu{\setlength{\unitlength}{0.00625in}%
\begin{picture}(215,174)(165,600)
\thicklines
\put(180,700){\line( 0,-1){ 40}}
\put(180,660){\line( 1, 0){ 40}}
\put(220,660){\line( 0, 1){ 40}}
\put(220,700){\line(-1, 0){ 40}}
\put(220,700){\line( 1, 2){ 20}}
\put(240,740){\line( 2, 1){ 40}}
\put(280,760){\line( 1, 0){ 40}}
\put(320,760){\line( 2,-1){ 40}}
\put(360,740){\line( 1,-2){ 20}}
\put(380,700){\line( 0,-1){ 40}}
\put(380,660){\line(-1,-2){ 20}}
\put(360,620){\line(-2,-1){ 40}}
\put(320,600){\line(-1, 0){ 40}}
\put(280,600){\line(-2, 1){ 40}}
\put(240,620){\line(-1, 2){ 20}}
\put(165,660){\makebox(0,0)[lb]{\raisebox{0pt}[0pt][0pt]{\tenrm 1}}}
\put(165,700){\makebox(0,0)[lb]{\raisebox{0pt}[0pt][0pt]{\tenrm 2}}}
\put(210,705){\makebox(0,0)[lb]{\raisebox{0pt}[0pt][0pt]{\tenrm 3}}}
\put(230,740){\makebox(0,0)[lb]{\raisebox{0pt}[0pt][0pt]{\tenrm 4}}}
\put(270,765){\makebox(0,0)[lb]{\raisebox{0pt}[0pt][0pt]{\tenrm 5}}}
\put(230,610){\makebox(0,0)[lb]{\raisebox{0pt}[0pt][0pt]{\tenrm n-1}}}
\put(215,650){\makebox(0,0)[lb]{\raisebox{0pt}[0pt][0pt]{\tenrm n}}}
\end{picture}
}

\def\figdia{\setlength{\unitlength}{0.00625in}%
\begin{picture}(215,160)(165,600)
\thicklines
\put(220,700){\line( 1, 2){ 20}}
\put(240,740){\line( 2, 1){ 40}}
\put(280,760){\line( 1, 0){ 40}}
\put(320,760){\line( 2,-1){ 40}}
\put(360,740){\line( 1,-2){ 20}}
\put(380,700){\line( 0,-1){ 40}}
\put(380,660){\line(-1,-2){ 20}}
\put(360,620){\line(-2,-1){ 40}}
\put(320,600){\line(-1, 0){ 40}}
\put(280,600){\line(-2, 1){ 40}}
\put(240,620){\line(-1, 2){ 20}}
\put(180,680){\line( 2, 1){ 40}}
\put(220,700){\line( 2,-1){ 40}}
\put(260,680){\line(-2,-1){ 40}}
\put(180,680){\line( 2,-1){ 40}}
\put(165,680){\makebox(0,0)[lb]{\raisebox{0pt}[0pt][0pt]{\tenrm 1}}}
\put(205,710){\makebox(0,0)[lb]{\raisebox{0pt}[0pt][0pt]{\tenrm 2}}}
\put(230,750){\makebox(0,0)[lb]{\raisebox{0pt}[0pt][0pt]{\tenrm 3}}}
\put(255,685){\makebox(0,0)[lb]{\raisebox{0pt}[0pt][0pt]{\tenrm n}}}
\put(195,655){\makebox(0,0)[lb]{\raisebox{0pt}[0pt][0pt]{\tenrm n-1}}}
\put(220,615){\makebox(0,0)[lb]{\raisebox{0pt}[0pt][0pt]{\tenrm n-2}}}
\end{picture}
}

\begin{de}
The graph $\Theta^n_{p,q} (q < p\leq n-1 < 2p+q)$ has
$n$ nodes labelled $\qset(n)$, and bonds,
$$\bonds({\Theta^n_{p,q}}):=
\{\begin{array}{l}(1,n),\,(p+q,n),\, (p,p+q+1),\, \\
(i\;i+1); i=\qset(n-1)\setminus \{p+q\}.
\end{array}\}.$$

\end{de}

\noindent (Se figure 6.)
Note that $\Theta^n_{n-1,0}=\ahat(n)$.

\begin{pr}
All unsplitting graphs $G$ that are not closed chain graphs
contain $\Theta^n_{p,q}$ as a subgraph for some positive
integers $p,q$.

\end{pr}

\centerline{
\setlength{\unitlength}{0.00625in}%
\begin{picture}(340,339)(165,390)
\thicklines
\put(420,700){\line(-1, 0){161}}
\put(259,699){\line(-3,-5){ 82.941}}
\put(179,559){\line( 3,-5){ 82.765}}
\put(260,420){\line( 1, 0){161}}
\put(421,421){\line( 3, 5){ 82.941}}
\put(501,561){\line(-3, 5){ 82.765}}
\put(180,560){\line( 1, 0){320}}
\put(200,600){\makebox(0,0)[lb]{\raisebox{0pt}[0pt][0pt]{$\bullet$}}}
\put(220,630){\makebox(0,0)[lb]{\raisebox{0pt}[0pt][0pt]{$\bullet$}}}
\put(240,665){\makebox(0,0)[lb]{\raisebox{0pt}[0pt][0pt]{$\bullet$}}}
\put(255,695){\makebox(0,0)[lb]{\raisebox{0pt}[0pt][0pt]{$\bullet$}}}
\put(295,695){\makebox(0,0)[lb]{\raisebox{0pt}[0pt][0pt]{$\bullet$}}}
\put(335,695){\makebox(0,0)[lb]{\raisebox{0pt}[0pt][0pt]{$\bullet$}}}
\put(375,695){\makebox(0,0)[lb]{\raisebox{0pt}[0pt][0pt]{$\bullet$}}}
\put(415,695){\makebox(0,0)[lb]{\raisebox{0pt}[0pt][0pt]{$\bullet$}}}
\put(435,665){\makebox(0,0)[lb]{\raisebox{0pt}[0pt][0pt]{$\bullet$}}}
\put(455,635){\makebox(0,0)[lb]{\raisebox{0pt}[0pt][0pt]{$\bullet$}}}
\put(470,600){\makebox(0,0)[lb]{\raisebox{0pt}[0pt][0pt]{$\bullet$}}}
\put(495,560){\makebox(0,0)[lb]{\raisebox{0pt}[0pt][0pt]{$\bullet$}}}
\put(475,520){\makebox(0,0)[lb]{\raisebox{0pt}[0pt][0pt]{$\bullet$}}}
\put(180,560){\makebox(0,0)[lb]{\raisebox{0pt}[0pt][0pt]{$\bullet$}}}
\put(195,520){\makebox(0,0)[lb]{\raisebox{0pt}[0pt][0pt]{$\bullet$}}}
\put(215,490){\makebox(0,0)[lb]{\raisebox{0pt}[0pt][0pt]{$\bullet$}}}
\put(255,420){\makebox(0,0)[lb]{\raisebox{0pt}[0pt][0pt]{$\bullet$}}}
\put(295,420){\makebox(0,0)[lb]{\raisebox{0pt}[0pt][0pt]{$\bullet$}}}
\put(335,420){\makebox(0,0)[lb]{\raisebox{0pt}[0pt][0pt]{$\bullet$}}}
\put(375,420){\makebox(0,0)[lb]{\raisebox{0pt}[0pt][0pt]{$\bullet$}}}
\put(415,420){\makebox(0,0)[lb]{\raisebox{0pt}[0pt][0pt]{$\bullet$}}}
\put(455,480){\makebox(0,0)[lb]{\raisebox{0pt}[0pt][0pt]{$\bullet$}}}
\put(435,450){\makebox(0,0)[lb]{\raisebox{0pt}[0pt][0pt]{$\bullet$}}}
\put(235,455){\makebox(0,0)[lb]{\raisebox{0pt}[0pt][0pt]{$\bullet$}}}
\put(185,600){\makebox(0,0)[lb]{\raisebox{0pt}[0pt][0pt]{\tenrm 1}}}
\put(210,640){\makebox(0,0)[lb]{\raisebox{0pt}[0pt][0pt]{\tenrm 2}}}
\put(225,675){\makebox(0,0)[lb]{\raisebox{0pt}[0pt][0pt]{\tenrm 3}}}
\put(305,720){\makebox(0,0)[lb]{\raisebox{0pt}[0pt][0pt]{$\cdots$}}}
\put(480,610){\makebox(0,0)[lb]{\raisebox{0pt}[0pt][0pt]{\tenrm p-1}}}
\put(505,565){\makebox(0,0)[lb]{\raisebox{0pt}[0pt][0pt]{\tenrm p}}}
\put(485,520){\makebox(0,0)[lb]{\raisebox{0pt}[0pt][0pt]{\tenrm p+q+1}}}
\put(165,560){\makebox(0,0)[lb]{\raisebox{0pt}[0pt][0pt]{\tenrm n}}}
\put(170,520){\makebox(0,0)[lb]{\raisebox{0pt}[0pt][0pt]{\tenrm n-1}}}
\put(180,485){\makebox(0,0)[lb]{\raisebox{0pt}[0pt][0pt]{\tenrm n-2}}}
\put(465,480){\makebox(0,0)[lb]{\raisebox{0pt}[0pt][0pt]{\tenrm p+q+2}}}
\put(255,560){\makebox(0,0)[lb]{\raisebox{0pt}[0pt][0pt]{$\bullet$}}}
\put(215,560){\makebox(0,0)[lb]{\raisebox{0pt}[0pt][0pt]{$\bullet$}}}
\put(295,560){\makebox(0,0)[lb]{\raisebox{0pt}[0pt][0pt]{$\bullet$}}}
\put(335,560){\makebox(0,0)[lb]{\raisebox{0pt}[0pt][0pt]{$\bullet$}}}
\put(375,560){\makebox(0,0)[lb]{\raisebox{0pt}[0pt][0pt]{$\bullet$}}}
\put(415,560){\makebox(0,0)[lb]{\raisebox{0pt}[0pt][0pt]{$\bullet$}}}
\put(455,560){\makebox(0,0)[lb]{\raisebox{0pt}[0pt][0pt]{$\bullet$}}}
\put(445,575){\makebox(0,0)[lb]{\raisebox{0pt}[0pt][0pt]{\tenrm p+1}}}
\put(210,575){\makebox(0,0)[lb]{\raisebox{0pt}[0pt][0pt]{\tenrm p+q}}}
\put(305,580){\makebox(0,0)[lb]{\raisebox{0pt}[0pt][0pt]{$\cdots$}}}
\put(300,390){\makebox(0,0)[lb]{\raisebox{0pt}[0pt][0pt]{$\cdots$}}}
\end{picture}
}

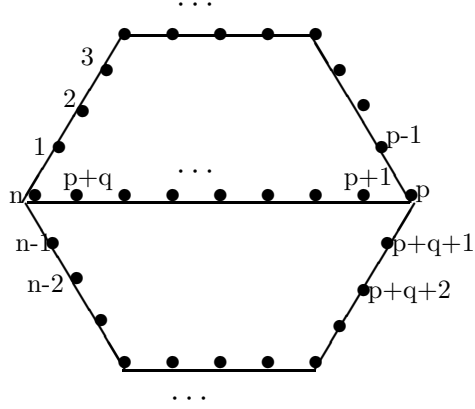
\noindent {\bf Figure 6} {\tenrm The graph $\Theta^n_{p,q}$ defined
above.}

%\medskip

%\medskip
%\centerline{\figtri\hfil\figsqu\hfil\figdia}
%\smallskip

\begin{pr}

For $G=\Theta^n_{p,q}$, let $\beta_1:=\{p+q+1,p+q+2,
\ldots,n-1\}$, $\beta_2:=\{1,2,\ldots,p-1\}$ and
$\beta_3:=\{p+1,p+2,\ldots,p+q\}$, and let
$\alpha_i:=\qset(n)\setminus\beta_i,\, i=1,2,3$.  Then,
for $a_i\in\alpha_i, i=1,2,3$, and defining
$\Phi_{\alpha_i\{a_i\}}^{\{a_i\}}\subseteq
\Phi_{\{a_i\}}^{\{a_i\}}$ to be the words of the
form $\omega_{A_{\alpha_i}}$ as in eq. (\ref{move}), we
have
$\Phi_{\alpha_1\{a_1\}}^{\{a_1\}}\cong{\sf Z}_{p+q}$,
$\Phi_{\alpha_2\{a_2\}}^{\{a_2\}}
\cong{\sf Z}_{n-p}$  and
$\Phi_{\alpha_3\{a_3\}}^{\{a_3\}}\cong{\sf Z}_{n-q-1}$.
Also $\varphi_{\{a_i\}}^{\{a_i\}}(x)=x$,
$\forall \varphi_{\{a_i\}}^{\{a_i\}}
\in\Phi_{\alpha_i\{a_i\}}^{\{a_i\}}$ and $x\in\beta_i$ for
$i=1,2,3$.

\end{pr}

\begin{co}
For $G=\Theta_{p,q}^n$, $\alpha\subset\nodes(G),
|\alpha|\geq 1$, $\exists
\varphi_{\alpha}^{\alpha}\in\Phi_{\alpha}^{\alpha}$
such that for $x,y\in\nodes(G)\setminus\alpha$,
$\varphi_{\alpha}^{\alpha}(x)=1$ and
$\varphi_{\alpha}^{\alpha}(y)=n$.
\end{co}
\noindent{\em Proof}:  This is achieved by a sequence of words as in the
proposition that ``moves" one of $x$, $y$, while keeping the
other fixed.  Such a move is easiest if $x\in\alpha_i$ and $y\in\alpha_j$,
$i\neq j$.  If $x,y\in\alpha_i$, we first move both until
only one of $x,y$ is in $\beta_j$ for some $j$.
{\hfill $\Box$}

\begin{co}
For $\Theta^n_{p,q}$ as above, and $n\geq 4$,
\ba
H_{\tri}^{n-1}=&\sym(n-1),\cr
H_{\squ}^{n-1}=&\cases{\sym(n-1)\; \quad\hbox{for}\quad
 \,n\; \hbox{odd}\cr
\alt(n-1)\;\quad\hbox{for}\quad \,n\,\;\hbox{even}.}\cr
H_{\dia}^{n-1}=&\cases{\alt(n-1)\; \quad\hbox{for}\quad
 \,n\; \hbox{odd}\cr
\sym(n-1)\;\quad\hbox{for}\quad\,n\,\;\hbox{even}.}
\cr
\ea
\end{co}

\noindent{\em Proof:}  By the
same procedure as in the corollary above,
it is possible to construct $\varphi_{\{n\}}^{\{n\}}
\in\Phi_{\{n\}}^{\{n\}}$, such that for

\noindent  $G=\tri$,
$\varphi_{\{n\}}^{\{n\}}(x)=n-2$ and $\varphi_{\{n\}}^{\{n\}}(y)=n-1$ for
$x,y\in\nodes(G)
\setminus \{n\}$ and

\noindent  $G=\squ,\dia$, $\varphi_{\{n\}}^{\{n\}}(x_i)
=n-i,i=1,2,3$  and $x_i\in\nodes(G) \setminus \{n\}$.

Thus, for $\tri$ we can achieve arbitrary transpositions,
which generate $\sym(n-1)$, and for $\squ, \dia$ we can
achieve all $3$-cycles, generating $\alt(n-1)$.  However,
by the proposition above, we can also realize
$\Phi_{\alpha_1\{a_1\}}^{\{a_1\}}\cong{\sf Z}_{p+q}$,
which for $G=\squ, \dia$ are ${\sf Z}_{n-3}$ and
${\sf Z}_{n-2}$ respectively.  For $n$ even (odd)
these would give even (odd) permutations for $\squ, \dia$
respectively. Hence the result.
{\hfill $\Box$}

\end{subsection}

\begin{subsection}{On constructing a partition basis
for $\Dd(G)$.}

In any partition in $\Ss(2n)$, perform the operations of
ignoring either the elements $i'$ or the elements
$\bt(i)$ for $i\in\qset(n)$.  These may be viewed as
sub-partitions of the nodes $\bt(i)$ and $i'$,
which we call ``bottoms'' and ``tops'' respectively.
The elements of $\Db$, the partition basis of $\Dg(G)$ can be constructed
diagrammatically, by
figuring out the possible ``top" and ``bottom" configurations
that can be achieved by drawing connectivities on $G\times
A_k$ for $k$ large, and the possible ways of gluing
the top and bottom by the $\Hh(z)=i$ ``propagating lines."
The ways of joining bottom to top are dictated by $H_G^i$,
so the next step in this program is to determine the set of
allowed tops, the bottoms being isomorphic under up-down
transposition.

\begin{pr}
For $G=\Theta^n_{p,q}, \; (q\leq p<n-1<2p+q)$,
a partition basis
element $z\in\Delta_i\; (i<n)$ is also in the
partition basis $\Bb(i)^G$ of a $\Dg(G)\Ee(i)$-module
iff

\noindent i)$\exists$
at least one part of $z$ of the form $(a')$
(a singleton node, $a\in\nodes(G)$), or

\noindent ii) in the sub-partition of $z$ consisting of
nodes $i'$, $i\in\nodes(G)$, one of the parts
is of the form
$(\cdots a'b'\cdots)$,
where $(a,b)\in\bonds(G)$,

\noindent and the rest may be partitioned
in any arbitrary way.

\noindent For all elements of the form i), the configurations of ``tops'' in
$\Bb(i)^G$ depend only on
$n$ and not on $p,q$.

\end{pr}

\noindent {\em Proof}:

(Only if.) Every word except $1$ in $\Dd(G)$ must begin with
either $\Ai(j),j\in\nodes(G)$ or
$\Aij(i,j), (i,j)\in\bonds(G)$.

(If.) (By construction of such a $z$.)
Without loss of generality,
let the partition $z$ have the primed nodes in a
sub-partition of the
shape $(l_1,l_2,\ldots,l_r,1)$ where the last part is the
 singleton $(a')$ as in case i).  For case ii), all
words may be written as $\Aij(a,b) z$, with $z$ constructed
as in case i).
For $k=1,2,\ldots i$,
the parts are
$(a^{(k)}_1 \; a^{(k)}_2 \; \ldots \;
a^{(k)}_{l_k}\pi(\bt(k)))$,
where $a^{(k)}_j,\,j=1,\ldots,l_k$ are the primed nodes
and $\pi(\bt(k))$ is the image (under $\pi\in H_G^i$)
of the bottom node $\bt(k)$.
For parts numbered $k=i+1,i+2,\ldots,r$, the parts are
of the form $(a^{(k)}_1 \; a^{(k)}_2 \;
\ldots \; a^{(k)}_{l_k})$,
the $(r+1)^{th}$ part is the singleton $(a')$,
and the remaining parts consist of singleton
bottom nodes.

For all $\alpha,\beta\in\nodes(G)$
such that $|\alpha|=j=|\beta|$, let
$\overline{\Phi}_j:=\cup_{\alpha,\beta}
\Phi_{\alpha}^{\beta}$.  For a given $z$,
define the words
$$\varphi_j^{(k)}\in\overline{\Phi}_{s(z,j,k)}, \;(1\leq k\leq r,
1\leq j\leq l_k-1) \; \;  \mbox{where}   \; \;
s(z,j,k)=\sum_{u=1}^{k-1}l_u+j-\mbox{min}(i,k-1),$$
and $\psi_j \in \overline{\Phi}_{l_j}$, by
\beq
\begin{array}{lccll}
\varphi_1^{(k)}(a_1) & = &n, & & 1\leq k\leq r \\
\varphi_j^{(k)}(a_{j+1}^{(k)})&=&1,& \forall j=1,2,\ldots l_k-1, & 1\leq k \leq
r\\
\varphi_j^{(k)}(1)&=&n & \forall j=2,3,\ldots, l_k-1,
& 1\leq k \leq r\\
\varphi_j^{(k)}(\pi(j))&=&\pi(\bt(j))& & k>j \\
\psi_k(1)&=&\pi(\bt(k))&& k=1,2,\ldots, i,
\end{array}
\eq
where the domain and ranges of the elements of
$\overline{\Phi}_{s(z,j,k)},\;\overline{\Phi}_{l_j}$
are nodes of $G$.  Note that, $\pi(\bt(k))$ indicates the
positions of the bottom nodes in $z$ as required.
Such a construction is possible
by corollary 4.1.
The word
$$\Ai(a)
\Biggl(\prod_{k=1}^{i}\biggl[ (\prod_{j=1}^{l_k-1} \varphi_j^{(k)}
\Aij(1,n)\Ai(n))\psi_k \biggr]
\prod_{k=i+1}^r\biggl[ (\prod_{j=1}^{l_k-1} \varphi_j^{(k)}
\Aij(1,n)\Ai(n) ) \Ai(1)\biggr]\Biggr)\Ee(i)$$
in  $\Dd(G)$ constructs the required $z$.  (See figure 7).
{\hfill$\Box$}

\medskip

\centerline{
\setlength{\unitlength}{0.00900in}%
\begin{picture}(300,599)(80,210)
\thicklines
\put( 80,630){\dashbox{4}(260,20){}}
\put( 80,600){\dashbox{4}(260,20){}}
\put( 80,500){\dashbox{4}(260,20){}}
\put( 80,470){\dashbox{4}(260,20){}}
\put(100,530){\line( 0,-1){ 70}}
\put(120,530){\line( 0,-1){ 70}}
\put(140,530){\line( 0,-1){ 70}}
\put(160,530){\line( 0,-1){ 70}}
\put(180,530){\line( 0,-1){ 70}}
\put(200,530){\line( 0,-1){ 70}}
\put(220,530){\line( 0,-1){ 70}}
\put(240,530){\line( 0,-1){ 70}}
\put(260,530){\line( 0,-1){ 70}}
\put(280,530){\line( 0,-1){ 70}}
\put(320,530){\line( 0,-1){ 70}}
\put(300,530){\line( 0,-1){ 40}}
\put(300,470){\line( 0,-1){ 10}}
\put(300,510){\line( 1, 0){ 20}}
\put( 80,370){\dashbox{4}(260,20){}}
\put( 80,340){\dashbox{4}(260,20){}}
\put(100,400){\line( 0,-1){ 70}}
\put(120,400){\line( 0,-1){ 70}}
\put(140,400){\line( 0,-1){ 70}}
\put(160,400){\line( 0,-1){ 70}}
\put(180,400){\line( 0,-1){ 70}}
\put(200,400){\line( 0,-1){ 70}}
\put(220,400){\line( 0,-1){ 70}}
\put(240,400){\line( 0,-1){ 70}}
\put(260,400){\line( 0,-1){ 70}}
\put(280,400){\line( 0,-1){ 70}}
\put(320,400){\line( 0,-1){ 70}}
\put(300,400){\line( 0,-1){ 40}}
\put(300,340){\line( 0,-1){ 10}}
\put(300,380){\line( 1, 0){ 20}}
\put( 80,740){\dashbox{4}(260,20){}}
\put( 80,660){\framebox(260,60){}}
\put( 80,530){\framebox(260,60){}}
\put( 80,400){\framebox(260,60){}}
\put( 80,270){\framebox(260,60){}}
\put(140,780){\line( 0,-1){ 20}}
\put(100,780){\line( 0,-1){ 60}}
\put(100,720){\line( 4,-3){ 80}}
\put(180,660){\line( 0,-1){200}}
\put(180,460){\line(-2,-3){ 40}}
\put(140,400){\line( 0,-1){ 10}}
\put(120,780){\line( 0,-1){ 60}}
\put(120,720){\line( 4,-3){ 80}}
\put(200,660){\line( 0,-1){ 70}}
\put(200,590){\line( 1,-1){ 60}}
\put(260,530){\line( 0,-1){ 70}}
\put(320,780){\line( 0,-1){ 60}}
\put(320,720){\line(-4,-3){ 80}}
\put(240,660){\line( 0,-1){ 70}}
\put(240,590){\line(-4,-3){ 80}}
\put(160,530){\line( 0,-1){200}}
\put(160,330){\line( 2,-3){ 40}}
\put(200,270){\line( 0,-1){ 30}}
\put(160,780){\line( 0,-1){ 60}}
\put(160,720){\line(-1,-1){ 60}}
\put(100,660){\line( 0,-1){ 70}}
\put(100,590){\line( 1,-3){ 20}}
\put(120,530){\line( 0,-1){200}}
\put(120,330){\line( 2,-3){ 40}}
\put(160,270){\line( 0,-1){ 30}}
\put(140,390){\line( 0,-1){150}}
\put(120,650){\line( 0,-1){ 20}}
\put(140,650){\line( 0,-1){ 20}}
\put(160,650){\line( 0,-1){ 20}}
\put(260,650){\line( 0,-1){ 20}}
\put(280,650){\line( 0,-1){ 20}}
\put(300,650){\line( 0,-1){ 20}}
\put(320,650){\line( 0,-1){ 20}}
\put(300,640){\line( 1, 0){ 20}}
\put(120,630){\line( 0,-1){ 40}}
\put(140,630){\line( 0,-1){ 40}}
\put(160,630){\line( 0,-1){ 40}}
\put(260,630){\line( 0,-1){ 40}}
\put(280,630){\line( 0,-1){ 40}}
\put(320,630){\line( 0,-1){ 40}}
\put(300,620){\line( 0, 1){ 10}}
\put(300,590){\line( 0, 1){ 10}}
\put(240,780){\line( 0,-1){ 60}}
\put(240,720){\line( 4,-3){ 80}}
\put(320,660){\line( 0,-1){ 10}}
\put(280,780){\line( 0,-1){ 60}}
\put(280,720){\line( 1,-3){ 20}}
\put(300,660){\line( 0,-1){ 10}}
\put(140,740){\line( 0,-1){ 20}}
\put(220,780){\line( 0,-1){ 60}}
\put(220,720){\line( 2,-3){ 40}}
\put(260,660){\line( 0,-1){ 10}}
\put(180,780){\line( 0,-1){ 60}}
\put(180,720){\line(-1,-1){ 60}}
\put(120,660){\line( 0,-1){ 10}}
\put(140,650){\line( 0, 1){ 10}}
\put(140,660){\line( 1, 1){ 60}}
\put(200,720){\line( 0, 1){ 60}}
\put(300,780){\line( 0,-1){ 60}}
\put(300,720){\line(-1,-3){ 20}}
\put(280,660){\line( 0,-1){ 10}}
\put(160,650){\line( 0, 1){ 10}}
\put(160,660){\line( 5, 3){100}}
\put(260,720){\line( 0, 1){ 60}}
\put(220,660){\line( 0,-1){ 70}}
\put(320,590){\line(-1,-3){ 20}}
\put(320,460){\line(-1,-3){ 20}}
\put(260,590){\line( 1,-1){ 60}}
\put(280,590){\line( 0,-1){ 60}}
\put(160,590){\line( 2,-3){ 40}}
\put(120,590){\line( 1,-3){ 20}}
\put(140,590){\line( 4,-3){ 80}}
\put(140,460){\line(-2,-3){ 40}}
\put(280,460){\line(-4,-3){ 80}}
\put(220,460){\line( 2,-3){ 40}}
\put(260,460){\line(-1,-3){ 20}}
\put(200,460){\line( 2,-1){120}}
\put(100,330){\line( 0,-1){ 90}}
\put(260,330){\line( 1,-1){ 60}}
\put(320,270){\line( 0,-1){ 30}}
\put(200,330){\line(-1,-3){ 20}}
\put(180,270){\line( 0,-1){ 30}}
\put(240,330){\line( 1,-1){ 60}}
\put(300,270){\line( 0,-1){ 30}}
\put(320,330){\line(-1,-1){ 60}}
\put(260,270){\line( 0,-1){ 30}}
\put(380,740){\makebox(0,0)[lb]{\raisebox{0pt}[0pt][0pt]{$\Ai(a), \, a=4$}}}
\put(380,700){\makebox(0,0)[lb]{\raisebox{0pt}[0pt][0pt]{$\varphi_1^{(1)}$}}}
\put(380,640){\makebox(0,0)[lb]{\raisebox{0pt}[0pt][0pt]{$\Aij(n,1), n=12$}}}
\put(380,600){\makebox(0,0)[lb]{\raisebox{0pt}[0pt][0pt]{$\Ai(n)$}}}
\put(380,560){\makebox(0,0)[lb]{\raisebox{0pt}[0pt][0pt]{$\varphi_2^{(1)}$}}}
\put(380,500){\makebox(0,0)[lb]{\raisebox{0pt}[0pt][0pt]{$\Aij(n,1)$}}}
\put(380,470){\makebox(0,0)[lb]{\raisebox{0pt}[0pt][0pt]{$\Ai(n)$}}}
\put(380,430){\makebox(0,0)[lb]{\raisebox{0pt}[0pt][0pt]{$\varphi_3^{(1)}$}}}
\put(380,380){\makebox(0,0)[lb]{\raisebox{0pt}[0pt][0pt]{$\Aij(n,1)$}}}
\put(380,340){\makebox(0,0)[lb]{\raisebox{0pt}[0pt][0pt]{$\Ai(n)$}}}
\put(380,300){\makebox(0,0)[lb]{\raisebox{0pt}[0pt][0pt]{$\psi_1\circ
\varphi_1^{(2)}$}}}
\put(315,240){\makebox(0,0)[lb]{\raisebox{0pt}[0pt][0pt]{$\bullet$}}}
\put(295,240){\makebox(0,0)[lb]{\raisebox{0pt}[0pt][0pt]{$\bullet$}}}
\put(275,240){\makebox(0,0)[lb]{\raisebox{0pt}[0pt][0pt]{$\bullet$}}}
\put(255,240){\makebox(0,0)[lb]{\raisebox{0pt}[0pt][0pt]{$\bullet$}}}
\put(235,240){\makebox(0,0)[lb]{\raisebox{0pt}[0pt][0pt]{$\bullet$}}}
\put(215,240){\makebox(0,0)[lb]{\raisebox{0pt}[0pt][0pt]{$\bullet$}}}
\put(195,240){\makebox(0,0)[lb]{\raisebox{0pt}[0pt][0pt]{$\bullet$}}}
\put(175,240){\makebox(0,0)[lb]{\raisebox{0pt}[0pt][0pt]{$\bullet$}}}
\put(155,240){\makebox(0,0)[lb]{\raisebox{0pt}[0pt][0pt]{$\bullet$}}}
\put(135,240){\makebox(0,0)[lb]{\raisebox{0pt}[0pt][0pt]{$\bullet$}}}
\put(115,240){\makebox(0,0)[lb]{\raisebox{0pt}[0pt][0pt]{$\bullet$}}}
\put( 95,240){\makebox(0,0)[lb]{\raisebox{0pt}[0pt][0pt]{$\bullet$}}}
\put( 90,210){\makebox(0,0)[lb]{\raisebox{0pt}[0pt][0pt]{\tenrm \c{2}}}}
\put(110,210){\makebox(0,0)[lb]{\raisebox{0pt}[0pt][0pt]{\tenrm \c{3}}}}
\put(130,210){\makebox(0,0)[lb]{\raisebox{0pt}[0pt][0pt]{\tenrm \c{4}}}}
\put(150,210){\makebox(0,0)[lb]{\raisebox{0pt}[0pt][0pt]{\tenrm \c{5}}}}
\put(170,210){\makebox(0,0)[lb]{\raisebox{0pt}[0pt][0pt]{\tenrm \c{6}}}}
\put(190,210){\makebox(0,0)[lb]{\raisebox{0pt}[0pt][0pt]{\tenrm \c{7}}}}
\put(210,210){\makebox(0,0)[lb]{\raisebox{0pt}[0pt][0pt]{\tenrm \c{8}}}}
\put(230,210){\makebox(0,0)[lb]{\raisebox{0pt}[0pt][0pt]{\tenrm \c{9}}}}
\put(250,210){\makebox(0,0)[lb]{\raisebox{0pt}[0pt][0pt]{\tenrm \c{10}}}}
\put(270,210){\makebox(0,0)[lb]{\raisebox{0pt}[0pt][0pt]{\tenrm \c{11}}}}
\put(290,210){\makebox(0,0)[lb]{\raisebox{0pt}[0pt][0pt]{\tenrm \c{12}}}}
\put(315,210){\makebox(0,0)[lb]{\raisebox{0pt}[0pt][0pt]{\tenrm \c{1}}}}
\put( 95,780){\makebox(0,0)[lb]{\raisebox{0pt}[0pt][0pt]{$\bullet$}}}
\put(115,780){\makebox(0,0)[lb]{\raisebox{0pt}[0pt][0pt]{$\bullet$}}}
\put(135,780){\makebox(0,0)[lb]{\raisebox{0pt}[0pt][0pt]{$\bullet$}}}
\put(155,780){\makebox(0,0)[lb]{\raisebox{0pt}[0pt][0pt]{$\bullet$}}}
\put(175,780){\makebox(0,0)[lb]{\raisebox{0pt}[0pt][0pt]{$\bullet$}}}
\put(195,780){\makebox(0,0)[lb]{\raisebox{0pt}[0pt][0pt]{$\bullet$}}}
\put(215,780){\makebox(0,0)[lb]{\raisebox{0pt}[0pt][0pt]{$\bullet$}}}
\put(235,780){\makebox(0,0)[lb]{\raisebox{0pt}[0pt][0pt]{$\bullet$}}}
\put(255,780){\makebox(0,0)[lb]{\raisebox{0pt}[0pt][0pt]{$\bullet$}}}
\put(275,780){\makebox(0,0)[lb]{\raisebox{0pt}[0pt][0pt]{$\bullet$}}}
\put(295,780){\makebox(0,0)[lb]{\raisebox{0pt}[0pt][0pt]{$\bullet$}}}
\put(315,780){\makebox(0,0)[lb]{\raisebox{0pt}[0pt][0pt]{$\bullet$}}}
\put(110,800){\makebox(0,0)[lb]{\raisebox{0pt}[0pt][0pt]{\tenrm 3'}}}
\put(130,800){\makebox(0,0)[lb]{\raisebox{0pt}[0pt][0pt]{\tenrm 4'}}}
\put(150,800){\makebox(0,0)[lb]{\raisebox{0pt}[0pt][0pt]{\tenrm 5'}}}
\put(170,800){\makebox(0,0)[lb]{\raisebox{0pt}[0pt][0pt]{\tenrm 6'}}}
\put(190,800){\makebox(0,0)[lb]{\raisebox{0pt}[0pt][0pt]{\tenrm 7'}}}
\put(210,800){\makebox(0,0)[lb]{\raisebox{0pt}[0pt][0pt]{\tenrm 8'}}}
\put(230,800){\makebox(0,0)[lb]{\raisebox{0pt}[0pt][0pt]{\tenrm 9'}}}
\put(250,800){\makebox(0,0)[lb]{\raisebox{0pt}[0pt][0pt]{\tenrm 10'}}}
\put(270,800){\makebox(0,0)[lb]{\raisebox{0pt}[0pt][0pt]{\tenrm 11'}}}
\put(290,800){\makebox(0,0)[lb]{\raisebox{0pt}[0pt][0pt]{\tenrm 12'}}}
\put( 95,800){\makebox(0,0)[lb]{\raisebox{0pt}[0pt][0pt]{\tenrm 2'}}}
\put(315,800){\makebox(0,0)[lb]{\raisebox{0pt}[0pt][0pt]{\tenrm 1'}}}
\end{picture}
}

\medskip

\noindent{\bf Figure 7} {\tenrm
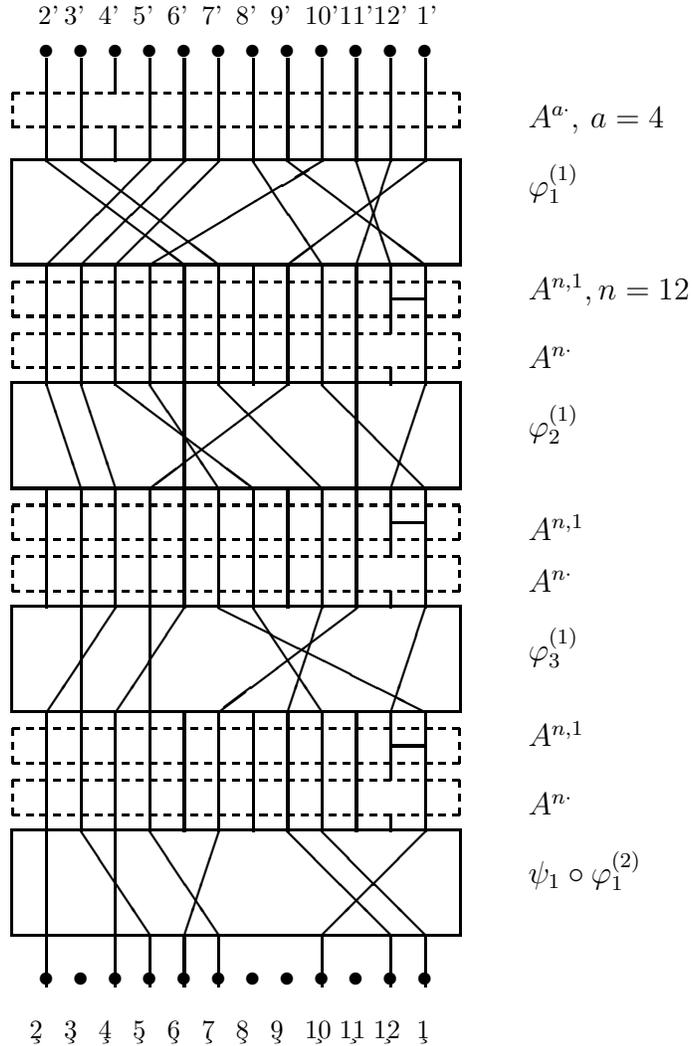
The figure depicts how the word $\Ai(4)$
($\prod_{j=1}^3 \varphi_j^{(1)} \Aij(12,1) \Ai(12)$)
$\psi_1$ builds the partition ((8$'$ 9$'$
10$'$ 11$'$ \c{3})(12$'$ \c{6})
(7$'$ \c{1})(6$'$ \c{2})(5$'$ \c{5})(3$'$ \c{12})(2$'$ \c{4})(1$'$
\c{7})(4$'$)(\c{8})(\c{9})(\c{10})(\c{11}) ) on the lattice
G $\times A_k$, where G is of the type $\Theta_{p,q}^{n=12}$.
The dashed boxes denote specific letters in $\Dd(G)$ while
those drawn with continuous lines are words in $D_G$.
Note that the number of lines in the interior of
each of the solid boxes above decreases from top to bottom.
}

\smallskip

This proposition will then enable us to estimate the
cardinality of the partition basis of $\Dg(G)$,
which we need
for the next sub-section.

\end{subsection}

\begin{subsection}{On locating the exceptional values of Q:}

The next stage in giving the  generic structure is to give the
dimensions of the irreducibles and an explicit construction of a basis for
each.
Let $\ep(\gamma)$ be a primitive idempotent of $H_G^i$,
i.e. such that
\[
R_{\gamma} = H_G^i \ep(\gamma),\quad \gamma\in\Gamma_G^i
\]
is an irreducible representation of $H^i_G$ (all of these are
known, from \cite{c-r} for example).
Then
\[
W_{\gamma} = \Dd(G) (\Ee(i)\otimes \ep(\gamma))
\]
builds generic irreducible $W_{\gamma}$.

In fact our present concern is not to determine the generic irreducible
dimensions for
a given $G$,
but to locate the exceptional $Q$ values (by analogy with
the Beraha numbers for $G=A_n$ which is relevant for the
two dimensional case). To this end it is highly indicative to proceed as
follows.

We first decide on a sequence approaching the large graph limit
(not the same as the thermodynamic limit, see below).
Thus, we take a sequence of graphs
\[
G^{(-)} = \{ G^{j} \; : \; j=1,2,.... \}
\]
(with, say, $A_l \times A_m$ ($l,m$ large) at the `end'
if we were to consider graphs appropriate for cubic lattice Potts models as in
the next section)
and then determine
\[
k^{i,\gamma}_{G^{(-)}} = \lim_{j \rightarrow \infty}
{{dim(\Dd(G^{(j)}) \Ee(i)^{(n_j)} \ep(\gamma))}
\over {dim(\Dd(G^{(j-1)}) \Ee(i)^{(n_{j-1})} \ep(\gamma))}}, \; \;
n_j=|\nodes({G^{(j)}})|,
\]
if it exists.

Since for the $Q$-state Potts model representation,
\[
{{dim(\pot(G^{(j)}) )}\over{ dim(\pot(G^{(j-1)}) )}} = Q^m
\]
for any sequence such that
$|\nodes(G^{(j)})|=|\nodes(G^{(j-1)})|+m$, $m$ a positive integer,
if $k^{i,\gamma}_{G^{(-)}} >Q^m$,
 then $Q$ is exceptional by the following
argument \cite{Mar}.
Since $\Ee(0)$ is a primitive idempotent
(i.e., $\Ee(0)\Dd(G)\Ee(0)=\C\Ee(0)$), $\Dd(G)\Ee(0)$
is indecomposable.  If $\Dg(G)$ were semisimple, $\Dd(G)\Ee(0)$ would be
contained in $\pot(G)$
with multiplicity 1 for all $G$.  Thus, the
evaluation of the case $\kappa_G:=k^{0,(0)}_{G^{(-)}}$
is sufficient for any sequence $G^{(-)}$.  So, if dim($\Dd(G)\Ee(0)$)$>Q^n$
(for $Q$ an integer) for
$n=|\nodes(G)|$, $Q$ is exceptional.
For example $k^{i,\gamma}_{A^{(-)}}=4$ where
$A^{(-)}=\{ G^{j}=A_j\}_{j\geq 1}$, and this signals the
special nature of the $Q=1,2,3$ state Potts models in two dimensions.

We can now consider the asymptotic growth rate of dimensions
of the irreducible representations.  In particular, we
estimate a lower bound on the dimension of the module $\Dd(G)\Ee(0)$.

\begin{pr}

For $G=\Theta^n_{p,q}, (q\leq p\leq n-2)$, and any
$k\in\Re$, there exists a natural number $M$, such that
dim($\Dd(G)\Ee(0)$) $ > k^n$ for  $n>M$.
\end{pr}

\noindent{\em Proof}:

Let $b_m$ be the number of ways of
pairing $m$ nodes (for some even number $m$).
For any node (say $1$), its partner (in the partition of
shape $(2^{m/2})$) can be chosen in $m-1$ ways, while
the rest of the pairs can be chosen in $b_{m-2}$ ways.
This determines
$b_m=(m-1)b_{m-2}$ for all even $m$, with $b_1=1$.
Thus for any $k\in\Re,\exists M$, a natural number, such
that $b_m>k^m$ for $m>M$.  From proposition 5, for
any basis element of $\Dd(G)$ for $G=\Theta^n_{p,q}$, $n-1$
primed nodes may be partitioned in any arbitrary way.
The number of such possibilities is clearly
larger that $b_{n-1}$
(where we have chosen $n$ to be odd, without loss of generality).
Therefore,
\[
\mbox{dim (}D_G \Ee(0)\mbox{)} >  k^n.
\]
{\hfill $\Box$}

Thus, for any integer $Q$ and $G$ of the type
 $\Theta_{p,q}^n$,
the dimension of $\Dd(G)\Ee(0)$ is larger than that of the
Potts representation, $\pot(G)$ ($>Q^n$ for $Q$ a positive integer).

We thus have

\begin{pr}
Consider a sequence $G^{(-)}:=\{G^{(j)}\}_{j\geq 1}$
of unsplitting graphs (except $G^{(j)}=\ahat(j)$).
For any positive integer $Q$, $\exists$ an
integer $n_1$ s.t. $\forall n_2>n_1$, $D_{G^{(n_2)}}$
is not semi-simple.

\end{pr}

\end{subsection}

\section{Cubic lattice Potts models: G=$A_l\times A_m$}

This is the case we are most interested in, to which we
shall apply the results obtained in section 2.

\begin{pr}  For $G=A_l\times A_m$, ($l,m\geq2$),
and $n=lm$,

\noindent i) $H_G^{n-1}=\alt(n-1)$\hfill

\noindent ii) $H_G^i=\sym(i), i<n-1$.

\end{pr}

\noindent {\em Proof}:
i) For $n$ even, $G$ has a subgraph $\squ$, and for $n$
odd, it has a subgraph $\dia$. Recall corollary 4.2.
ii) It is easy to see that $H_{\squ}^{n-2}$ and
$H_{\dia}^{n-2}$ for $n=4$ are isomorphic to each other and
to $H_{\tri}^{n-1}$ for $n=3$.
For $i\leq n-3$, recall (\ref{ron}).
{\hfill $\Box$}

\def\lbar{\overline{\lambda}}

Recall that the representations of $\alt(j)$ are indexed
by unordered pairs of partitions, $\lambda\vdash j$ and its conjugate
$\lambda'$, for $\lambda\neq\lambda'$.  For $\lambda
=\lambda'=\lbar$,
there are two non-isomorphic outer automorphism-conjugate  representations
labeled $\lbar$ and
$\lbar^{*}$.

\begin{co} For $G=A_l\times A_m$,
\[
\Gamma_G=\cup_{i=0}^{n-2}\{\lambda\vdash i\}
\cup\{(\lambda,\lambda'),\lbar,\lbar^{*}|
\lambda,\lbar\vdash n-1\}\cup \{\lambda\vdash n=(n)\}.
\]
is the index set for irreducible representations of
$\Dg(G)$.
\end{co}

Note that by filling in one of the diagonals of
an elementary plaquette of $G=A_l
\times A_m$, we obtain a graph which has $\tri$ as a subgraph.
The index set for the irreducible representations
of the diagram algebras for such graphs with $\tri$
subgraphs is the same as that of the complete graph on
$n$ nodes. (See \cite{marsal}.)

Since $A_l\times A_m\supset \Theta^n_{p,q}$ for
some $p$, $q$, we have (from proposition 5),

\begin{co}

For $G=A_l\times A_m, (n=lm)$,
a partition basis
element $z\in\Delta_i\; (i<n-1)$ is also in the
partition basis $\Bb(i)^G$ of a $\Dg(G)\Ee(i)$-module
iff

\noindent i)$\exists$
at least one part of $z$ of the form $(a')$
(a singleton node, $a\in\nodes(G)$), or

\noindent ii) in the sub-partition of $z$ consisting of
nodes $i'$, $i\in\nodes(G)$, one of the parts
is of the form
$(\cdots a'b'\cdots)$,
where $(a,b)\in\bonds(G)$,

\noindent and the rest may be partitioned
in any arbitrary way.

\end{co}

Also, since $G=A_l\times A_m$ is of the unsplitting type,
we infer

\begin{pr}

Consider a sequence $G^{(-)}:=\{G^{(j)}\}_{j\in {\sf Z}_{+} }$,
where $G^{(j)}=A_l\times A_m, l,m>1,lm=j$.
For any positive integer $Q$, $\exists$ an
integer $n_1$ s.t. $\forall n_2>n_1$, $D_{G^{(n_2)}}$
is not semi-simple.

\end{pr}

\def\word{
\setlength{\unitlength}{0.00825in}%
\begin{picture}(280,554)(220,260)
\thicklines
\put(240,800){\line( 0,-1){ 40}}
\put(260,800){\line( 0, 1){  0}}
\put(260,800){\line( 0,-1){ 40}}
\put(280,800){\line( 0,-1){ 40}}
\put(300,800){\line( 0, 1){  0}}
\put(300,800){\line( 0,-1){ 40}}
\put(320,800){\line( 0,-1){ 40}}
\put(340,800){\line( 0, 1){  0}}
\put(340,800){\line( 0,-1){ 40}}
\put(360,800){\line( 0,-1){ 40}}
\put(380,800){\line( 0, 1){  0}}
\put(380,800){\line( 0,-1){ 40}}
\put(240,720){\line( 0,-1){ 40}}
\put(260,720){\line( 0, 1){  0}}
\put(260,720){\line( 0,-1){ 40}}
\put(280,720){\line( 0,-1){ 40}}
\put(300,720){\line( 0, 1){  0}}
\put(300,720){\line( 0,-1){ 40}}
\put(320,720){\line( 0,-1){ 40}}
\put(340,720){\line( 0, 1){  0}}
\put(340,720){\line( 0,-1){ 40}}
\put(360,720){\line( 0,-1){ 40}}
\put(380,720){\line( 0, 1){  0}}
\put(380,720){\line( 0,-1){ 40}}
\put(240,640){\line( 0,-1){ 40}}
\put(260,640){\line( 0, 1){  0}}
\put(260,640){\line( 0,-1){ 40}}
\put(280,640){\line( 0,-1){ 40}}
\put(300,640){\line( 0, 1){  0}}
\put(300,640){\line( 0,-1){ 40}}
\put(320,640){\line( 0,-1){ 40}}
\put(340,640){\line( 0, 1){  0}}
\put(340,640){\line( 0,-1){ 40}}
\put(360,640){\line( 0,-1){ 40}}
\put(380,640){\line( 0, 1){  0}}
\put(380,640){\line( 0,-1){ 40}}
\put(220,720){\framebox(200,40){}}
\put(400,800){\line( 0,-1){ 40}}
\put(220,640){\framebox(180,40){}}
\put(360,620){\line( 1, 0){ 20}}
\put(240,600){\line( 0,-1){ 40}}
\put(260,600){\line( 0, 1){  0}}
\put(260,600){\line( 0,-1){ 40}}
\put(280,600){\line( 0,-1){ 40}}
\put(300,600){\line( 0, 1){  0}}
\put(300,600){\line( 0,-1){ 40}}
\put(320,600){\line( 0,-1){ 40}}
\put(340,600){\line( 0, 1){  0}}
\put(340,600){\line( 0,-1){ 40}}
\put(360,600){\line( 0,-1){ 40}}
\put(220,520){\framebox(160,40){}}
\put(240,560){\line( 3,-2){ 60}}
\put(280,560){\line(-1,-1){ 40}}
\put(260,560){\line( 5,-2){100}}
\put(360,560){\line(-2,-1){ 80}}
\put(260,520){\line( 1, 1){ 40}}
\put(340,560){\line( 0,-1){ 40}}
\put(320,560){\line( 0,-1){ 40}}
\put(240,520){\line( 0,-1){ 40}}
\put(260,520){\line( 0, 1){  0}}
\put(260,520){\line( 0,-1){ 40}}
\put(280,520){\line( 0,-1){ 40}}
\put(300,520){\line( 0, 1){  0}}
\put(300,520){\line( 0,-1){ 40}}
\put(320,520){\line( 0,-1){ 40}}
\put(340,520){\line( 0, 1){  0}}
\put(340,520){\line( 0,-1){ 40}}
\put(360,520){\line( 0,-1){ 40}}
\put(240,480){\line( 1, 0){ 40}}
\put(300,480){\line( 1, 0){ 20}}
\put(240,480){\line( 0,-1){ 40}}
\put(260,480){\line( 0, 1){  0}}
\put(260,480){\line( 0,-1){ 40}}
\put(280,480){\line( 0,-1){ 40}}
\put(300,480){\line( 0, 1){  0}}
\put(300,480){\line( 0,-1){ 40}}
\put(320,480){\line( 0,-1){ 40}}
\put(340,480){\line( 0, 1){  0}}
\put(340,480){\line( 0,-1){ 40}}
\put(360,480){\line( 0,-1){ 40}}
\put(240,440){\line( 0,-1){ 40}}
\put(300,440){\line( 0, 1){  0}}
\put(300,440){\line( 0,-1){ 40}}
\put(340,440){\line( 0, 1){  0}}
\put(340,440){\line( 0,-1){ 40}}
\put(360,440){\line( 0,-1){ 40}}
\put(220,360){\framebox(160,40){}}
\put(240,360){\line( 0,-1){ 20}}
\put(260,360){\line( 0,-1){ 20}}
\put(280,360){\line( 0,-1){ 20}}
\put(300,360){\line( 0,-1){ 20}}
\put(220,300){\framebox(100,40){}}
\put(240,300){\line( 0,-1){ 40}}
\put(260,300){\line( 0,-1){ 40}}
\put(240,400){\line( 0,-1){ 40}}
\put(260,360){\line( 1, 1){ 40}}
\put(280,360){\line( 3, 2){ 60}}
\put(300,360){\line( 3, 2){ 60}}
\put(240,680){\line( 3,-2){ 60}}
\put(260,680){\line(-1,-2){ 20}}
\put(280,680){\line(-1,-2){ 20}}
\put(300,680){\line( 1,-1){ 40}}
\put(320,680){\line( 3,-2){ 60}}
\put(340,680){\line( 1,-2){ 20}}
\put(360,680){\line(-1,-1){ 40}}
\put(380,680){\line(-5,-2){100}}
\put(235,800){\makebox(0,0)[lb]{\raisebox{0pt}[0pt][0pt]{$1$}}}
\put(255,800){\makebox(0,0)[lb]{\raisebox{0pt}[0pt][0pt]{$2$}}}
\put(395,800){\makebox(0,0)[lb]{\raisebox{0pt}[0pt][0pt]{$n$}}}
\put(305,805){\makebox(0,0)[lb]{\raisebox{0pt}[0pt][0pt]{$\ldots$}}}
\put(500,710){\makebox(0,0)[lb]{\raisebox{0pt}[0pt][0pt]{$\Ai(n)$}}}
\put(500,660){\makebox(0,0)[lb]{\raisebox{0pt}[0pt][0pt]{$\pi_e\in\alt(n-1)$}}}
\put(500,620){\makebox(0,0)[lb]{\raisebox{0pt}[0pt][0pt]{$\Aij(n-2,n-1)$}}}
\put(500,580){\makebox(0,0)[lb]{\raisebox{0pt}[0pt][0pt]{$\Ai(n-1)$}}}
\put(500,540){\makebox(0,0)[lb]{\raisebox{0pt}[0pt][0pt]{$\pi\in\sym(n-2)$}}}
\put(500,480){\makebox(0,0)[lb]{\raisebox{0pt}[0pt][0pt]{$\prod_{(r,s)\in
L}$}}}
\put(500,435){\makebox(0,0)[lb]{\raisebox{0pt}[0pt][0pt]{$\prod_{i\in
X}\Ai(i)$}}}
\put(500,385){\makebox(0,0)[lb]{\raisebox{0pt}[0pt][0pt]{$\omega\in$$
eq.(\ref{move})}}}
\put(500,320){\makebox(0,0)[lb]{\raisebox{0pt}[0pt][0pt]{$\prod_{j=i+1}^{n-2}
(\Ai(j)/Q)$}}}
\end{picture}
}
%}

%\medskip

%\noindent{\bf Figure 7} {\tenrm Constructing $z$ as in the %proof of
%Proposition 6.}

%\noindent {\bf Remark.} The necessary and sufficient %conditions for the
%%allowed 'tops' (primed nodes) is the
%same for all unsplitting graphs $G$, only the assignment
%of the unprimed nodes are dictated by the appropriate
%$H_G^i$.  This can be easily proved by a variant of the
%proof of proposition 6.

%By the same arguments
%as before, the growth rate (with $n=|\nodes(G)|$
%of the dimensions of $\Dd(G)\Ee(0)$ for such classes of
%graphs is always larger than that of the Potts
%representation, $\pot(G)$.
%In fact, the remark following proposition 6
%implies that this is true of all $\Dd(G)$ for
%unsplitting graphs $G$ excluding closed chain graphs.

\section{Discussion}

For $G=A_l\times
A_2$ for instance, it might have naively been expected that
for large $l$, the results of the familiar $A_l$ case
might be approached.  However, instead of the
known growth rate of dimensions, i.e. $4$, we get an
unboundedly large number.
This discrepancy might be attributed
to the length, $k$ of the graph $A_k$
in the transfer (``time-like") direction
of the lattice $L=G\times A_k$, on which the
partition function is evaluated.  The connectivities
of nodes are achieved involve
``permuting the nodes," i.e.,
the action of (\ref{phi}), where each shift (\ref{move}) can
only be realized for $k>\mu_i$.
A restriction of the
maximum $k$ allowed will obviously reduce the dimensions and
their growth rate $\kappa_G$.  This is clearly necessary to
define the true thermodynamic limit, where the volume has to
increase in a specified fashion, keeping the ratios of
lengths in all the directions of the lattice, $L$
fixed to some finite value, unlike in the
definition of $\kappa_G$, where
the size of the $G$ was increased independent of $k$.
In the two-dimensional
case, the connectivities can all be achieved on $L=A_n
\times A_k$ for $k\sim n$, and the problem does not
arise.  Thus, it might be useful to
define a certain ``cutoff" height $k$ of the
representations of
$\Dg(G)$ to narrow in on the physically relevant sectors of
the representation theory.

We have indicated that for
the smallest deviations away from chain graphs, e.g.,
$\tri$, the diagram algebra is too large to carry
directly useful physical information.  Suitable
quotients have to be implemented to
reduce the size of the representations and an
appropriately quotiented algebra would then be the
analogous ``generic" algebra for the cubic lattice
models.
The special values of $Q$ for which the algebra ceases to be
semi-simple is the obvious place to look for the quotient
relations that are relevant for the Potts representation
which is defined for integer values of $Q$.
These integers are certainly a subset of the special
points where the algebra ceases to be
semi-simple, as we have shown.
We expect that the techniques outlined
in the appendix can be extended to obtain the degeneracies of the cubic lattice
Potts spectrum,
which we would like to report
in the future.  We have also undertaken preliminary
calculations on the location of other $Q$-values for which
$\Dg(G)$ becomes non-semi-simple for $G=A_l\times A_m$,
and so far found only rationals.  Further studies are in progress.

\medskip

\noindent{\bf Acknowledgements}

The authors would like to thank the EPSRC for financial support in the
form of the grants GRJ25758 and GRJ29923.  P.M. would also like to thank
the Nuffield Foundation for partial financial support.

\section{Appendix}

\begin{pr} As left $\Dg(G)$ modules, $P_n\Ee(i)\cong
\Dd(G)\Ee(i)\oplus R^{G}_i$
modulo $\Dd(G)\Ee(i-1)\Dd(G)$
for any $G$, $|G|=n$, where $R^G_i$ is either empty or
a direct sum of $\theta_i$ copies of the
trivial $H_G^i$-modules,  $\chi_{(i)}$, $(i)\vdash i$,
$R^G_i\cong\theta_i\chi_{(i)}$.  Further, for
$\lambda\vdash i\in\{0,1,2,\ldots,n\},\,n=|\nodes(G)|$,
\eql(pd)
\res(P,G,D,G)\;\vgl=\bigoplus
\cases{
{}_{D_G} W_{\lambda} \oplus \theta_i \chi_{(i)} ,
\quad\forall i\in\qset(n-2)\cr
	\cases{
{}_{D_G} W_{\lambda,\lambda'} \oplus\theta_i \chi_{(i)},
\quad \lambda,\lambda'\vdash i=n-1,  \cr
{}_{D_G} W_{\lambda}\oplus {}_{D_G} W_{\lambda}^{*}
\oplus\theta_i \chi_{(i)} \quad\mbox{for}\lambda=
\lambda'\vdash i=n-1}\cr
\oplus\theta_i \chi_{(i)}, \quad\mbox{for} \quad i=n.}
\eq

\end{pr}

\noindent{\em Proof}:
$\Ai(j) P_n \Ee(i)=\Aij(k,l) P_n \Ee(i)=0$
mod $\Dd(G)\Ee(i)$, therefore,$\Dd(G)x=1x$
mod $\Dd(G)\Ee(i)\forall x\in P_n\Ee(i)$. $H_G^i$ acts trivially.
For $\lambda\vdash n-1$, the labels of the representations of
$\Pn$ and $\Dg(G)$ are those of $\sym(n-1)$ and $\alt(n-1)$
respectively, and $\mbox{Res}_{\alt(n-1)}^{\sym(n-1)}$ must be
invoked.
{\hfill $\Box$}

To characterize the generic structure of the algebra
completely, it is necessary to determine the dimensions of
its irreducible representations.
Also, it is useful to characterize the
inclusion of algebras, while approaching the
large graph limit described above, in order to identify the
subspaces that carry the information relevant for a
physical interpretation.  A preliminary step would be to
determine how, for $H\subset G$, $\Dg(G)$-modules split up
as $\Dg(H)$-submodules.  Henceforth, we shall denote
a left $R$-module $M$ as ${}_RM$.

\begin{pr}
For $G=A_l\times A_m$ and $G\supset H=A_{l-1}\times A_m$.
$$\Dd(G)\Ee(i)\cong{\oplus}_{j=-m}^{m}(\Dd(H)\Ee(i+j)
\oplus R_{i+j}^{G,H}),$$
as left $\Dg(H)$ modules, where $R_{i+j}^{G,H}$ is
either empty or $\theta_{i+j} \chi_{(i+j)}$, where
$\theta_k$ is the multiplicity of the trivial
$H_H^k$-module, $\chi_{(k)},(k)\vdash k$.
\end{pr}

\noindent{\em Proof}: Let $\underline{m}:=\{n-m+1,n-m+2,
\ldots,n\}$ and $p_{\underline{m}}:=\qset(n-m)$.
If $w\in\Dd(G)\Ee(i)$ s.t. none of the parts of
$w$ is of the form $(\cdots k'\cdots l'\cdots)$ for
$k'\in p_{\underline{m}}$ and $l'\in \underline{m}$,
$\Dd(H)w\cong {\oplus}_{j=0}^m\Dd(H)\Ee(i-j)$.  Each summand
indexed by $j$ denotes the number of parts of $w$ which
contain only primed nodes.

Similarly, if the nodes
of $\underline{m}$ are in some $j\leq m$ parts with nodes
of $p_{\underline{m}}$, we get
$\Dd(H)w\cong {\oplus}_{j=0}^m\Dd(H)\Ee(i+m-j)$, where once
again, $j$ counts the number of parts of $w$ containing only
primed nodes.

As before, for $x\in\Dd(G)\Ee(i)$, $\Dd(H)x=1x$ mod $\Dd(G)
\Ee(i)$, and $H_G^i$ thus acts trivially.
{\hfill $\Box$}

Let
$\wgg$ denote an irreducible left $\Dg(G)$ module,
$\gamma\in \Gamma_G$.  We are interested in
the restriction $\res(D,G,D,H)\; \wgg$.
Note the following inclusion of algebras:

\[
\begin{array}{lcl}
\Dg(G) & \subset & P_{|\nodes(G)|} \\
\cup   & \mbox{} & \cup    \\
\Dg(H) & \subset & P_{|\nodes(H)|}
\end{array}
\]
and consider the corresponding restrictions of modules:
\ba
\res(P,G,P,H)\;\vgl=&\oplus_{\mu}\xi_{\lambda\mu}
 \;\vhm,\quad
\lambda\in{\cal L}_{|\nodes(G)|},\mu\in
{\cal L}_{|\nodes(H)|},
\cr
\res(P,G,D,G)\;\vgl=&\oplus_{\gamma}g_{\lambda\gamma}\;
 \wgg, \quad
\lambda\in{\cal L}_{|\nodes(G)|},\gamma\in{\Gamma}_{G},
\cr
\res(P,H,D,H)\;\vhm=&\oplus_{\eta}h_{\mu\eta}\; \whe,
 \quad \lambda\in{\cal L}_{|\nodes(H)|},\mu\in{\Gamma}_{H},
\cr
\res(D,G,D,H)\;\wgg=&\oplus_{\mu}m_{\gamma\eta}\; \whe,
 \quad \gamma\in{\Gamma}_{G},\eta\in{\Gamma}_{H},
\cr
\ea
where $P_G:=P_{|\nodes(G)|}$ and (recall) the index sets
labelling the irreducible representations of $\Pn$ and $\Dg(G)$ are
${\cal L}_n$ and $\Gamma_G$ respectively.

Let dim($\wgg$) be denoted $d_{\gamma}$ and dim($\whe$)
 $:=d_\eta$.  Since the representations have already been
assigned an index set, it is sufficient to determine the
inclusion matrix ${\cal M}$, whose entries are the
multiplicities $m_{\gamma\eta}$ in
$$d_{\gamma}=\sum_{\eta} m_{\gamma\eta} d_{\eta}$$
in order to complete the study of the generic irreducibles.

To obtain this, recall that
$\res(P,G,P,H)$ is known, i.e.,
the coefficients $\xi_{\lambda\mu}\in\Xi^m$,
where the inclusion matrix $\Xi$ encodes the restriction
information $P_{n-1}(Q)\subset \Pn$ has been given in
\cite{s2p} and $m=|\nodes(G)|-|\nodes(H)|$.  For a
(left) $\Pn$-module, $\,{}_{P_{n}} V_{\lambda}$,$\lambda
\vdash i$,
$${\mbox{Res}}^{P_n}_{P_{n-1}}\;
{}_{P_{n}} V_{\lambda}=\bigoplus\cases{
{}_{P_{n-1}} V_{\lambda'},\,
{i-1}\dashv\lambda'\lhd\lambda\cr
{}_{P_{n-1}} V_{\lambda}\oplus{}_{P_{n-1}} V_{\lambda'},\,
i\dashv\lambda'\rhd\lhd\lambda\cr
{}_{P_{n-1}} V_{\lambda'},\,
i+1\dashv\lambda'\rhd\lambda,}$$
where $\lambda\rhd\mu$ denotes the ``removal of a box" from
$\lambda$ to produce $\mu$, $\lambda\lhd\mu$, denotes
the ``addition of a box" to $\lambda$ to produce $\mu$, and
$\lambda\rhd\lhd\mu$ means that we first remove a box from
$\mu$ to obtain some $\nu$ (say), and then add a box to $\nu$ to obtain
$\lambda$.  Addition and removal of boxes
correspond to the induction and restriction rules for
symmetric group representations, called the Pieri
(or Littlewood-Richardson) rules.  Also note that in the
above, ${}_{P_{n-1}} V_{\lambda}\cong
{}_{P_{n}} V_{\lambda}$.

This is the key piece of information which, together
with proposition (\ref{pd}), will indicate the way
to obtain $m_{\gamma\eta}$.
Let us evaluate $\res(P,G,D,H)$ in two ways,
corresponding to the paths in the diagram indicating
the inclusion of algebras above (restrictions are
transitive).
$$\res(P,G,D,H)=\res(D,G,D,H)\res(P,G,D,G)=\res(P,H,D,H)
\res(P,G,P,H).$$
Thus, from one path we get,
\beq
\res(P,G,D,H)\;\vgl=\res(P,H,D,H)\biggl(\res(P,G,P,H)\;\vgl
\biggr)=\oplus_{\mu}\oplus_{\eta}\xi_{\lambda\mu}
h_{\mu\eta}\; \whe,
\eq
while from the other,
\beq
\res(P,G,D,H)\vgl=\res(D,G,D,H)\biggl(\res(P,G,D,G)\vgl
\biggr)=\oplus_{\gamma}\oplus_{\eta} g_{\lambda\gamma}
m_{\gamma\eta} \; \whe.
\eq

Let the inclusion matrices $\Sigma_i$ and $\Upsilon_i^j$ encode the restriction
information $\mbox{Res}_{\sym(i)}
^{\alt(i)}$ and $\mbox{Res}_{\alt(i)}^{\alt(j)}$, with
matrix elements $(\Sigma_i)_{a,b}=\varsigma_{a,b}^{(i)}$
 and $(\Upsilon_i^j)_{a,b}=\upsilon_{a,b}^{(i,j)}$
respectively.  Then, the restriction information between representations that
are not among the list of one-dimensional
representations ($\chi_{(i)}$), is extracted from the above.
$$\tilde{m}_{\gamma\eta}=\cases{\xi_{\gamma\eta}, \gamma\vdash
k<n-1,\; \eta\vdash l<n-m-1,\cr
\sum_{\mu}\xi_{\gamma\mu}\varsigma^{(n-m-1)}_{\mu\eta},
 \gamma\vdash k<n-1,\; \; \mu,\eta\vdash l=n-m-1  \cr
\upsilon_{\gamma\eta}^{(n-1,n-m-1)},\; \gamma\vdash n-1,\,
\eta\vdash n-m-1.}$$
In the above, we have used $\tilde{m}_{\gamma\eta}$
instead of ${m}_{\gamma\eta}$ to indicate that $\tilde{m}_{\gamma\eta}$
does not give the multiplicities $\theta_i$
of the one-dimensional representations, $\chi_{(i)}$.
The number of such $\chi_{(i)}$ is not known in general.
Diagrammatically their determination is a combinatorial
problem of enumerating the number of ``top" configurations
that are characterized by corollary 9.2.

We have constructed an algorithm for their
enumeration by using recurrence relations for
$G=A_l\times A_2$, but we have not been able to
solve it in closed form.  For arbitrary rectangular graphs,
the combinatorics is much more complicated.

\end{document}